\documentclass{elsarticle}
\usepackage{amsmath,amsthm,amssymb}
\usepackage{graphicx,tikz,epsfig,caption,musixtex,minibox,multirow,makecell}
\usepackage{pgfplots}
\usepackage{rotating}
\usepackage{relsize}
\tikzset{fontscale/.style = {font=\relsize{#1}}
    }
  
\usepackage{pdfpages}
\usepackage{blkarray}
\usetikzlibrary{decorations.markings}
\usetikzlibrary{calc}
\usepackage{ulem}%to show deleted sentences.

\usepackage[utf8]{inputenc}

 \usepackage{graphics}
 \usepackage{tikz}
 \usetikzlibrary{arrows}

\usepackage{amssymb}

\usepackage{ulem}
\usepackage{cancel}

\usepackage{kotex}

\newcommand{\figref}[1]{Figure~\ref{#1}}
\newcommand{\tabref}[1]{Table~\ref{#1}}

\theoremstyle{plain}
\newtheorem{theorem}{Theorem}[section]

\theoremstyle{definition}
\newtheorem*{definition}{Definition}

\theoremstyle{remark}
\newtheorem{remark}[theorem]{Remark}

\begin{document}

\begin{frontmatter}
%
%%% Title, authors and addresses
%
\title{Topological Data Analysis of Korean Music in Jeongganbo: A Cycle Structure}
\author{Mai Lan Tran$^{ab}$, Changbom Park$^c$, Jae-Hun Jung$^{ab}$* \\
{\it a Department of Mathematics, Pohang University of Science \& Technology, Pohang 37673, Korea\\
b POSTECH Mathematical Institute for Data Science (MINDS), Pohang University of Science \& Technology, Pohang 37673, Korea
}\\
{\it c School of Physics, Korea Institute for Advanced Study, Seoul 02455, Korea}
}
\cortext[cor1]{Corresponding author. Emails: mailantran@postech.ac.kr (Mai Lan Tran), cbp@kias.re.kr (Changbom Park), jung153@postech.ac.kr (Jae-Hun Jung)}
%
%
%
%%%%%%%%%%%----Journal of Mathematics and Music-----%%%%%%%%
%%\author{
%%\name{Mai Lan Tran \textsuperscript{ab}, Changbom Park\textsuperscript{c}, Jae-Hun Jung \textsuperscript{ab}$^{\ast}$\thanks{$^\ast$Corresponding author. Emails: mailantran@postech.ac.kr (Mai Lan Tran), cbp@kias.re.kr (Changbom Park), jung153@postech.ac.kr (Jae-Hun Jung)}}
%%\affil{\textsuperscript{a} Department of Mathematics, Pohang University of Science \& Technology, Pohang 37673, Korea; 
%%\textsuperscript{b} POSTECH Mathematical Institute for Data Science (MINDS), Pohang University of Science \& Technology, Pohang 37673, Korea;
%%\textsuperscript{c} School of Physics, Korea Institute for Advanced Study, Seoul 02455, Korea}
%%%\received{v6.0 released January 2015}
%%}
%%\maketitle
%%%%%%%%%%%----Journal of Mathematics and Music-----%%%%%%%%
%
%
%
\begin{abstract}
Jeongganbo is a unique music representation invented by Sejong the Great. Contrary to the western music notation, the pitch of each note is encrypted and the length is visualized directly in a matrix form in Jeongganbo. We use topological data analysis (TDA) to analyze the Korean music written in Jeongganbo for  
Suyeonjang, Songuyeo, and Taryong, those well-known pieces played at the palace and among noble community. We are particularly interested in the cycle structure. We first define and determine the node elements of each music, characterized uniquely with its pitch and length. Then we transform the music into a graph and define the distance between the nodes as their adjacent occurrence rate. The graph is used as a point cloud whose homological structure is investigated by measuring the hole structure in each dimension. We identify cycles of each music, match those in Jeongganbo, and show how those cycles are interconnected. The main discovery of this work is that the cycles of Suyeonjang and Songuyeo, categorized as a special type of cyclic music known as Dodeuri, frequently overlap each other when appearing in the music while the cycles found in Taryong, which does not belong to Dodeuri class, appear individually.

%We  provide the transformation relation between the graph node (each note). We also found that there exist some cycles that do not appear as a whole in Jeongganbo directly. Although they are not visible, we conjecture that those also play a crucial role in Jeongganbo music. 
\end{abstract}

%%%%%%%%%%%----Journal of Mathematics and Music-----%%%%%%%%
%%\begin{keywords}
%%Korean music; Jeongganbo music; Topological data analysis; Persistent homology; Cycles; Cyclic music 
%%%\textbf{(Please provide five to ten keywords taken from terms used in your manuscript)}
%%\end{keywords}
%%\begin{classcode}\textit{Classification codes}: AMS 00A65 \end{classcode}
%%%%%%%%%%%----Journal of Mathematics and Music-----%%%%%%%%

\begin{keyword}
Korean music; Jeongganbo music; Topological data analysis; Persistent homology; Cycles; Cyclic music 
%\textbf{(Please provide five to ten keywords taken from terms used in your manuscript)}
\end{keyword}
%
%%\textit{Classification codes}: AMS 00A65 
%
\end{frontmatter}

%%%%%%%%%%%%%
%\begin{center}
%{\Large{{Topological Data Analysis of Korean Music in Jeongganbo: An application to Suyenjang, Songuyeo and Taryong}}} \\
%\vspace{0.5cm}
%{\large{Tran Mai Lan, Jae-Hun Jung \\Department of Mathematics, Pohang University of Science and Technology 
%\\Changbom Park\\
%Korea Institute for Advanced Study
%} }\\%
%\end{center}
%%%%%%%%%%%
\section{Introduction}
Topological data analysis (TDA) via persistent homology provides an efficient way of analyzing the cycle or loop structures embedded in multi-dimensional data. Particularly the one-dimensional homology structure is closely related to the repeating patterns in music flow when it mapped to the proper topological space. We consider, in this research, three famous Korean old music pieces, namely {\color{black}{{\it Suyeonjangjigok} (Suyeonjang hereafter), {\it Songkuyeojigok} (Songkuyeo hereafter)}}, and {\it Taryong} and study their cycle structures when represented in a music network — the cycle structures are the key element that characterizes these music pieces.  They are categorized as “Jeong-Ak” meaning the ‘Right Music’ mostly played at the palace and among noble community and were written in Jeongganbo, a unique music representation invented by Sejong the Great of Joseon dynasty. Contrary to the western music notation, the pitch of each note is encrypted and the length is visualized directly in a matrix form of Jeongganbo. As {\color{black}{the music}} is written in Jeongganbo, it sometimes provides a special musical pattern according to the Jeongganbo's matrix form. In this paper, we study the cycle structures and how those cycles are interconnected over the music flows of {\it Suyeonjang} and {\it Songkuyeo} and compare them with those of {\it Taryong}. According to the earliest literature in 1451 (Korea-sa-ak-ji (高麗史樂志) volumes 70 \& 71) the primitive form of {\it Suyeonjang} and {\it Songkuyeo}  appeared before the Joseon dynasty (1392$\sim$1897) in Korea. Suyeonjang is {\color{black}{originally a Chinese feast dance}} played at the palace wishing the king's longevity. {\color{black}{It has later evolved throughout Joseon dynasty to an orchestral ensemble with variations, and became one of the most representative Korean Jeong-Ak musics}}. Songkuyeo is the music composed in a similar pattern as Suyeonjang but with one octave higher. Suyeonjang and Songkuyeo have their unique music structure known as {\color{black}{“Dodeuri”}}, which means repeat-and-return. As Songkuyeo is composed with one octave higher, it is also called as “{\color{black}{Wut-Dodeuri}}” (Upper {\color{black}{Dodeuri}}) as Suyeonjang referred as “Mit-{\color{black}{Dodeuri}}” (Lower {\color{black}{Dodeuri}}). 

The {\color{black}{"Dodeuri"}} is a unique music structure. Its simplest pattern has a form of A-B-C-B where the second pattern C-B is the variation of the first pattern of A-B while the second part B of each pattern is repeated in both patterns. While this {\color{black}{"Dodeuri"}} music is old and has been studied for long, research is rare that studies how to define repeating cycles, how these repeating cycles are interconnected over the music flow, and what characteristics of the {\color{black}{"Dodeuri"}} patterns make both Suyeonjang and Songkuyeo unique in Jeong-Ak. In this study, we use the notion of persistent homology to define cycles embedded in Suyeonjang and Songkuyeo after we represent them as a music network. In order to construct the music network of these music pieces, we first define and determine the node elements, that constitute the music, characterized uniquely with its pitch and length. Then we transform the music into a graph and introduce a proper metric to define the distance between the notes using their adjacent occurrence rate. If two nodes appear, side-by-side, frequently we consider that those nodes are close. The graph with the metric is then used as a point cloud. For the analysis, we compute persistent homology of the constructed point cloud, especially for the one-dimensional homology. Through this research we identified cycles from each music as the cycles identified through persistent homology. Among those cycles found, some match the music, i.e. appear in the music in their found form, while some do not appear in the music directly. Then, by visualizing the distribution of those cycles in the music, we showed how those cycles are interconnected in the music. The main discovery of this work is that the cycles of Suyeonjang and Songuyeo, defined using homology, frequently overlap each other when appearing over the music while the cycles found in Taryong, which does not belong to {\it {\color{black}{Dodeuri}}} class, appear rather individually when they appear in the music. This gives an interesting musical effect that the listener feels he or she hears multiple cycles played simultaneously when they appear in the music. 
%This is, in fact, the witness from the audiences who hear the music played. 

The outline of the paper is as follows. In Section \ref{sec:tda} we briefly explain TDA, particularly persistent homology, and how we associate our music data with TDA tools in order to study the cycle structure in the music. Section \ref{sec:jeongganbo} reviews basic characteristics of Korean musical notation Jeongganbo, in particular Suyeonjang, Songkuyeo and Taryong. In Section \ref{sec:construction} we construct music network from music data and define the distance matrix which will be used to generate barcode for each music by Javaplex \citep{ATVJ} with Vietoris-Rips method. Section \ref{sec:freq} provides the frequency of the occurrence of each node in music. It is observed that the frequency decays more exponentially than algebraically. Detailed analysis of the barcodes and comparison between Suyeonjang, Songkuyeo and Taryong is given in Section \ref{sec:analysis}. At the end of the paper we provide the list of all nodes of Suyeonjang, Songkuyeo and Taryong, according to our node definition, in Tables \ref{tab:syj_allnodes},  \ref{tab:songkuyeo_allnodes} and  \ref{tab:taryong_allnodes} in Appendix.

\section{Topological data analysis} \label{sec:tda}
%\section{Analysis of Suyeonjang using TDA}
In this section we briefly explain TDA, particularly persistent homology. For more detailed explanation we refer to \citep{Carlsson}. The main motivation of using TDA is to study the cycle structure in music data. In this paper, we focus on the one-dimensional cycle, which is reflected as a repeated pattern in music. The two-dimensional cycle (or a two-dimensional void) would be also of interest, but its interpretation is uncertain at the moment. Below, we explain singular homology and then persistent homology. In the end, the cycle structure is visualized in the so-called barcode. The barcode is the key tool used in this paper. The point cloud used for TDA is constructed in Section \ref{sec:construction}. 

To begin with, we first explain a simplex. Let $\sigma_k$ be the oriented $k$-simplex. The $k$-simplex, $\sigma_k = [x_0, x_1, \cdots, x_k]$ is the convex hull of $k+1$ geometrically independent points (or vertices), $\{x_0, x_1, \cdots, x_k \} \subset R^n$, $k \le n$. Roughly speaking $\sigma_n$ is an $n$ dimensional triangle.  For example, $\sigma_0$ is a point or vertex, $\sigma_1$ an edge, $\sigma_2$ a triangle and $\sigma_3$ a tetrahedron. A higher dimensional simplex is the higher dimensional equivalent to a triangle. 
In this paper, the independent point is replaced with the node in a music network in the following section. 
In the following, 0-simplex (vertex), 1-simplex (edge) and 2-simplex (triangle) are shown from left to right, respectively. Here note
that the inside of triangle (n-dimensional triangle) is filled. 

\begin{center}
\begin{tikzpicture}
\filldraw[color=blue!99, fill=blue!99, very thick](-1,0) circle (0.2);

\filldraw[color=blue!99, fill=blue!99, very thick](1,0) circle (0.2);
\filldraw[color=blue!99, fill=blue!99, very thick](3,0) circle (0.2);
\draw [color = blue!99] (1,0) -- (3,0); 

\path[draw, fill=green!20] (5,0)--(6,1.6)--(7,0)--cycle;
\filldraw[color=blue!99, fill=blue!99, very thick](5,0) circle (0.2);
\filldraw[color=blue!99, fill=blue!99, very thick](6,1.6) circle (0.2);
\filldraw[color=blue!99, fill=blue!99, very thick](7,0) circle (0.2);
\draw [color = blue!99] (5,0) -- (6,1.6); 
\draw [color = blue!99] (6,1.6) -- (7,0); 
\draw [color = blue!99] (7,0) -- (5,0); 

%
%
%\fill[blue!50] (2.5,0) ellipse (1.5 and 0.5);
%\draw[ultra thick, ->] (6.5,0) arc (0:220:1);
\end{tikzpicture}
\end{center}

%            \begin{center}\begin{tikzpicture}[scale = 1]
%                \tikzstyle{every node}=[draw,shape=circle,inner sep=0pt,minimum size=.6cm,inner sep=0pt,minimum size=.6cm];
%                \node[fill=blue!99] (v1) at (1,0) {\textcolor{white}{\includegraphics[width=0.36cm]{Imw.png}}};%{$6$};
%                \node[fill=blue!99] (v2) at (2,0) {\textcolor{white}{\includegraphics[width=0.36cm]{namw.png}}};%{$11$};
%                \node[fill=red!30] (v3) at (3,0) {\textcolor{black}{\includegraphics[width=0.36cm]{jung0.png}}};%{$2$};
%                \node[fill=orange!99] (v4) at (4,0) {\textcolor{white}{\includegraphics[width=0.36cm]{Imw.png}}};%{$7$};
%                \node[fill=teal!60] (v5) at (5,0) {\textcolor{black}{\includegraphics[width=0.36cm]{jung0.png}}};%{$0$};
%                \node[fill=blue!99] (v6) at (6,0) {\textcolor{white}{\includegraphics[width=0.36cm]{Imw.png}}};%{$6$};
%                \node[draw=none] (v01) at (1.1,-0.5)  \node[fill=blue!99];%{$n_{6}$};
%                \node[draw=none] (v02) at (2.1,-0.5) {$n_{11}$};
%                \node[draw=none] (v03) at (3.1,-0.5) {$n_{2}$};
%                \node[draw=none] (v04) at (4.1,-0.5) {$n_{7}$};
%                \node[draw=none] (v05) at (5.1,-0.5) {$n_{0}$};
%                \node[draw=none] (v06) at (6.1,-0.5) {$n_{6}$};
%                \draw (v1) -- (v2)
%                    (v2) -- (v3)
%                    (v3) -- (v4)
%                    (v4) -- (v5)
%                    (v5) -- (v6);
%            \end{tikzpicture}\end{center}  
%

To explain singular homology, 
consider a ring $R$ and a topological space $X$, $(X,R)$. $R$ can be replaced with {\color{black}{$\mathbb{Q}$}}. In our case, 
the topological space $X$ is the music network constructed from the Jeongganbo music. 
 %In our research, the given Jeongganbo data is more likely $Q$.  
 Given $(X,R)$, let $C_n(X, R)$ be the free $R$-module generated by all possible continuous images of $\sigma_n$.  Consider $\delta_n:$ $C_n \rightarrow C_{n-1}$, the boundary map defined as below
$$
\delta_n\sigma_n = \sum_{k=0}^n (-1)^k[p_0,\dots, p_{k-1}, p_{k+1}, \dots, p_n]
$$
where $\{p_i\}$ are the vertices, the music nodes in our case. 
Then it is easy to show that $\delta_{n} \circ \delta_{n+1} = 0$ for $\forall n = 0, 1, \cdots$. 
For $n = 0$, we choose $\delta_0$ as a trivial map. 
Using the relation of vanishing consecutive boundary maps, we consider two subgroups, the kernel and image groups, $ker(\delta_n):= Z_n$ and $im(\delta_{n+1}):= B_n$, respectively. The kernel group is also called the cycle group and the image group is also called the boundary group. Obviously $Z_n\subset C_n$ and  $B_n \subset C_n$. Further, $B_n  \subseteq Z_n$, i.e. every boundary group is a cycle group. 

Using $B_n  \subseteq Z_n$, the $n$th homology group of $X$ with the coefficient $R$ is defined by the following quotient group 
$$
H_n(X, R) = Z_n/ B_n = ker(\delta_n)/im(\delta_{n+1}).  
$$
\textcolor{black}{For a detailed, rigorous description of homology we refer the reader to \citep{Hatcher, Munkres}}. Practically the $n$-dimensional homology group indicates how many $n$-dimensional {\it holes} are there in $X$. For example, for the sphere $H_0 = Z^1, H_1 = Z^0, H_2 = Z^1$ while for a torus $H_0 = Z^1, H_1 = Z^2, H_2 = Z^1$. The existence of holes and the number of holes are useful information that we will use in this research. Some useful theorems of using  homology for the practical problem can be found in \citep{Bergomithesis, Cohen, EHarer}. 

It is, however, difficult to compute $H_n$ in general and we use the notion of  simplicial complex out of the given data to understand the hole structures. Roughly speaking, with the simplicial complex, we approximate the original topological space $X$. Once the simplicial complex is constructed, we find its homology. \textcolor{black}{Related articles on using topological tools in music analysis are found in \citep{Bergomi, Bigo} and references therein}.
%With this notion, we will construct a simplicial complex using simplices. In this research, we will use homology group of the constructed simplicial complex. 

The tuple composed of  pitch and length in Jeongganbo is regarded as a node and all the nodes in the music constitutes a point cloud, which we will use to construct the simplicial complex. The simplicial complex is a topological space constructed by gluing a finite number of simplices. Since the exact topological space $X$ is unknown, we use its approximate topological space, i.e. the simplicial complex. In this research, for the simplicial complex, we consider the Vietoris-Rips complex. The basic procedure is as follows. From the given point cloud, we create one-simplices by connecting vertices resulting in edges, then create two-simplices by connecting edges and repeat this procedure for the construction of higher simplices.  For the connecting condition, we rely on the notion of metric. That is, the topology is described as metric topology. For the metric, we need to introduce the proper metric. In many cases, Euclidean metric is used to measure the pair-wise distance between simplices. For our research, we will use a different definition of metric defined in the following section based on the occurrence of consecutive nodes in the music flow. 
The procedure of connecting simplices is conducted if the connecting condition is satisfied. 
For the connecting condition, let us introduce  an auxiliary parameter, so-called the filtration parameter, $\tau$. If the distance between two simplices is less than the given value of the filtration parameter, $\tau$, those two simplices will be connected. 
%In our research, we use the Euclidean metric for distance. Here we note that it is also possible to use different metrics rather than the Euclidean metric. But we only focus on the Euclidean metric for the current research.
Once such a simplicial complex is constructed, we compute homology group at a particular value of $d$. Then we {\it collect }all the homology group in the sequence of $\tau$. This provides the notion of persistent homology. 

We also use the $n$-dimensional Betti number, $\beta_n$. The $n$-dimensional Betti number is the number of generators of the $n$th homology group. That is, $\beta_n$ is the number of copies of $R$.  When $n = 0$, $\beta_0$ simply means the total number of connected components. For example, for a torus, {\color{black}{$\beta_0 = 1$, $\beta_1 = 2$ and $\beta_2 = 1$}}. Once the simplicial complex is constructed from the Jeongganbo data, we will measure the Betti number in each dimension.  One can compute the $n$th Betti number of the constructed simplicial complex at a particular value of $\tau$.  The barcode is the collection of the Betti number with respect to the given value of $\tau$ in each dimension. For example, if $\tau= 0$, no two vertices are to be connected and the $0$th Betti number is the number of all connected components, which is simply the total number of vertices, nodes in our case. And $n$th Betti number will be zero for $n \ge 1$. For computational purpose and data analysis, one can plot $\beta_n$ versus $\tau$, known as barcode. There exists a certain interval of $\tau$ where the homology invariant in $n$ dimension. Then in barcode, such interval will be visualized as a persistent line, which we will call a {\it persistence}. The starting value of $\tau$ and the ending value of $\tau$  of such line are 
 called the {\it birth} and {\it death}. The plot of the death versus birth is called the persistence diagram. 
 %We basically use these two graphical measures to compare different Jenogganbo music. 

%\begin{figure}[th]
%	\centering
%		\includegraphics[width=0.35\textwidth]{Rips.png}
%		\includegraphics[width=0.35\textwidth]{Cech.png}
%	     \caption{
%	     http://en.wikipedia.org/wiki/Vietoris-Rips\_complex. Left: Vietoris-Rips complex. Right: \v{C}ech complex. }
%		%$\frac{dE}{dt}$ for $\eta = 0.9805$. $\frac{dE}{dt}$ is increasing strictly. (fig:syj_numofcyc805)}
%	 \label{figure1}
%\end{figure}

For example, we consider the following simple data composed of four music nodes, each a vertex of a unit square. Then the minimum distance between two random nodes will be $1$ and the maximum distance will be $\sqrt{2}$, the diagonal distance between the orange and red nodes or the green and blue nodes. 
\begin{center}
\begin{tikzpicture}
%\filldraw[color=blue!99, fill=blue!99, very thick](-1,0) circle (0.2);
%
%\filldraw[color=blue!99, fill=blue!99, very thick](1,0) circle (0.2);
%\filldraw[color=blue!99, fill=blue!99, very thick](3,0) circle (0.2);
%\draw [color = blue!99] (1,0) -- (3,0); 
%
%\path[draw, fill=green!20] (5,0)--(6,1.6)--(7,0)--cycle;
\filldraw[color=blue!99, fill=blue!99, very thick](1,0) circle (0.2);
\filldraw[color=orange!99, fill=orange!99, very thick](2,0) circle (0.2);
\filldraw[color=teal!60, fill=teal!60, very thick](2,1) circle (0.2);
\filldraw[color=red!30, fill=red!30, very thick](1,1) circle (0.2);
%\draw [color = blue!99] (5,0) -- (6,1.6); 
%\draw [color = blue!99] (6,1.6) -- (7,0); 
%\draw [color = blue!99] (7,0) -- (5,0); %
%
%\fill[blue!50] (2.5,0) ellipse (1.5 and 0.5);
%\draw[ultra thick, ->] (6.5,0) arc (0:220:1);
\end{tikzpicture}
\end{center}
Now using various values of $\tau\ge 0$, let us connect nodes. Here note that if a triangle is formed as a result, the triangle should be filled. The following shows how these four nodes are connected with different value of $\tau = 0, 0.5, 1, \sqrt{2}, 2$. The Betti numbers $\beta_0$ and $\beta_1$ are also provided for the give $\tau$. For example, when $\tau_1 = 1$, all the nodes are connected, so there is just one connected component, that is, $\beta_0 = 1$. As shown in the figure, the connected nodes also form one hole, a square. This $\beta_1 = 1$. But when $\tau = \sqrt{2}$, the square becomes two triangles as the blue and green nodes are connected with $\tau = \sqrt{2}$ and the hole disappear as those two triangles are filled inside. Note that it is also possible to connect the red and orange nodes instead of connecting the blue and green nodes. But once one of them is established, the other one can not be established because the non-empty intersection of any two simplices is a simplex. 

\begin{center}
\begin{tikzpicture}
\filldraw[color=blue!99, fill=blue!99, very thick](1,0) circle (0.2);
\filldraw[color=orange!99, fill=orange!99, very thick](2,0) circle (0.2);
\filldraw[color=teal!60, fill=teal!60, very thick](2,1) circle (0.2);
\filldraw[color=red!30, fill=red!30, very thick](1,1) circle (0.2);
\node at (1.5, -1) (nodeXi) {{\bf $\tau = 0$}};
\node at (1.5, -2) (nodeXi) {$\beta_0 = 4$};
\node at (1.5, -2.5) (nodeXi) {$\beta_1 = 0$};

\filldraw[color=blue!99, fill=blue!99, very thick](4,0) circle (0.2);
\filldraw[color=orange!99, fill=orange!99, very thick](5,0) circle (0.2);
\filldraw[color=teal!60, fill=teal!60, very thick](5,1) circle (0.2);
\filldraw[color=red!30, fill=red!30, very thick](4,1) circle (0.2);
\node at (4.5, -1) (nodeXi) {$\tau = 0.5$};
\node at (4.5, -2) (nodeXi) {$\beta_0 = 4$};
\node at (4.5, -2.5) (nodeXi) {$\beta_1 = 0$};

%\draw [color = blue!99] (1,0) -- (3,0); 
%\path[draw, fill=green!20] (5,0)--(6,1.6)--(7,0)--cycle;
\draw [color = blue!99] (7,0) -- (8,0); 
\draw [color = blue!99] (7,0) -- (7,1); 
\draw [color = blue!99] (7,1) -- (8,1); 
\draw [color = blue!99] (8,1) -- (8,0); 
\filldraw[color=blue!99, fill=blue!99, very thick](7,0) circle (0.2);
\filldraw[color=orange!99, fill=orange!99, very thick](8,0) circle (0.2);
\filldraw[color=teal!60, fill=teal!60, very thick](8,1) circle (0.2);
\filldraw[color=red!30, fill=red!30, very thick](7,1) circle (0.2);
\node at (7.5, -1) (nodeXi) {$\tau = 1$};
\node at (7.5, -2) (nodeXi) {$\beta_0 = 1$};
\node at (7.5, -2.5) (nodeXi) {$\beta_1 = 1$};

\draw [color = blue!99] (10,0) -- (11,0); 
\draw [color = blue!99] (10,0) -- (10,1); 
\draw [color = blue!99] (10,1) -- (11,1); 
\draw [color = blue!99] (11,1) -- (11,0); 
\draw [color = blue!99] (10,0) -- (11,1); 
\path[draw, fill=green!20] (10,0)--(10,1)--(11,1)--cycle;
\path[draw, fill=red!20] (10,0)--(11,0)--(11,1)--cycle;
\filldraw[color=blue!99, fill=blue!99, very thick](10,0) circle (0.2);
\filldraw[color=orange!99, fill=orange!99, very thick](11,0) circle (0.2);
\filldraw[color=teal!60, fill=teal!60, very thick](11,1) circle (0.2);
\filldraw[color=red!30, fill=red!30, very thick](10,1) circle (0.2);
\node at (10.5, -1) (nodeXi) {$\tau = \sqrt{2}$};
\node at (10.5, -2) (nodeXi) {$\beta_0 = 1$};
\node at (10.5, -2.5) (nodeXi) {$\beta_1 = 0$};

\draw [color = blue!99] (13,0) -- (14,0); 
\draw [color = blue!99] (13,0) -- (13,1); 
\draw [color = blue!99] (13,1) -- (14,1); 
\draw [color = blue!99] (14,1) -- (14,0); 
\draw [color = blue!99] (13,0) -- (14,1); 
\path[draw, fill=green!20] (13,0)--(13,1)--(14,1)--cycle;
\path[draw, fill=red!20] (13,0)--(14,0)--(14,1)--cycle;
\filldraw[color=blue!99, fill=blue!99, very thick](13,0) circle (0.2);
\filldraw[color=orange!99, fill=orange!99, very thick](14,0) circle (0.2);
\filldraw[color=teal!60, fill=teal!60, very thick](14,1) circle (0.2);
\filldraw[color=red!30, fill=red!30, very thick](13,1) circle (0.2);
\node at (13.5, -1) (nodeXi) {$\tau = 2$};
\node at (13.5, -2) (nodeXi) {$\beta_0 = 1$};
\node at (13.5, -2.5) (nodeXi) {$\beta_1 = 0$};

\end{tikzpicture}
\end{center}

For the above example we used only $5$ values of $\tau$. If we use multiple values of $\tau$ as many as we want and  plot  $\beta_0$ and $\beta_1$ versus $\tau$,  the following barcodes are obtained. We plot the points at each $\tau$ as many as the given Betti number in the vertical axis. 
\begin{center}
\begin{tikzpicture}
%\filldraw[color=blue!99, fill=blue!99, very thick](1,0) circle (0.2);
%\filldraw[color=orange!99, fill=orange!99, very thick](2,0) circle (0.2);
%\filldraw[color=teal!60, fill=teal!60, very thick](2,1) circle (0.2);
%\filldraw[color=red!30, fill=red!30, very thick](1,1) circle (0.2);
%\node at (1.5, -1) (nodeXi) {{\bf $\tau = 0$}};
%\node at (1.5, -2) (nodeXi) {$\beta_0 = 4$};
%\node at (1.5, -2.5) (nodeXi) {$\beta_1 = 0$};

\draw [color = black!99] (1,0) -- (7,0); 
\draw [color = black!99] (1,0) -- (1,5); 

\draw [color = black!99, line width=3mm] (1,1) -- (3,1); 
\draw [color = black!99, line width=3mm] (1,2) -- (3,2); 
\draw [color = black!99, line width=3mm] (1,3) -- (3,3); 
\draw [color = black!99, line width=3mm] (1,4) -- (7,4); 

\draw [color = black!99] (3,-0.1) -- (3,0.1); 
\node at (3,-0.5) (nodeXi) {$\tau = 1$};
\node at (2,-1) (nodeXi) {$\mbox{Barcode for } H_0 $};

\node at (0.5,1) (nodeXi) {$1$};
\node at (0.5,2) (nodeXi) {$2$};
\node at (0.5,3) (nodeXi) {$3$};
\node at (0.5,4) (nodeXi) {$4$};

%\draw [color = black!99] (9,0) -- (16,0); 
\draw [color = black!99] (9,0) -- (15,0); 
\draw [color = black!99] (9,0) -- (9,5); 

\draw [color = black!99] (11,-0.1) -- (11,0.1); 
\node at (10.5,-0.5) (nodeXi) {$\tau = $};
\node at (11,-0.5) (nodeXi) {$1$};

\draw [color = black!99, line width=3mm] (11,1) -- (11.9,1); 
\draw [color = black!99] (11.9,-0.1) -- (11.9,0.1); 
\node at (11.9,-0.5) (nodeXi) {$\sqrt{2}$};
\node at (8.5,1) (nodeXi) {$1$};

\node at (10,-1) (nodeXi) {$\mbox{Barcode for } H_1$};

\end{tikzpicture}
\end{center}
For later use, we define the persistence $\Pi^n_i$ denoting the length of $i$th barcode in $n$th dimension. For example, 
for the above example, 
$$
  \Pi^0_1 = \Pi^0_2 = \Pi^0_3 = 1, 
$$
and 
$$
  \Pi^1_1 = \sqrt{2} - 1. 
$$
As $\tau \rightarrow \infty$, there remains only one connected component implying $\Pi^0_4 \rightarrow \infty$.

%As the Vietoris-Rips complex is expensive to construct, we use the Lazy-witness algorithm instead. The Lazy-witness algorithm uses only smaller number of sampling points from the given point cloud in order to reduce the computational complexity of the Vietoris-Rips complex. We compute the topological measures multiple times for the same point cloud. 

\section{Korean musical notation Jeongganbo} \label{sec:jeongganbo}

\begin{figure}[!h]\centering
\includegraphics[scale=0.6]{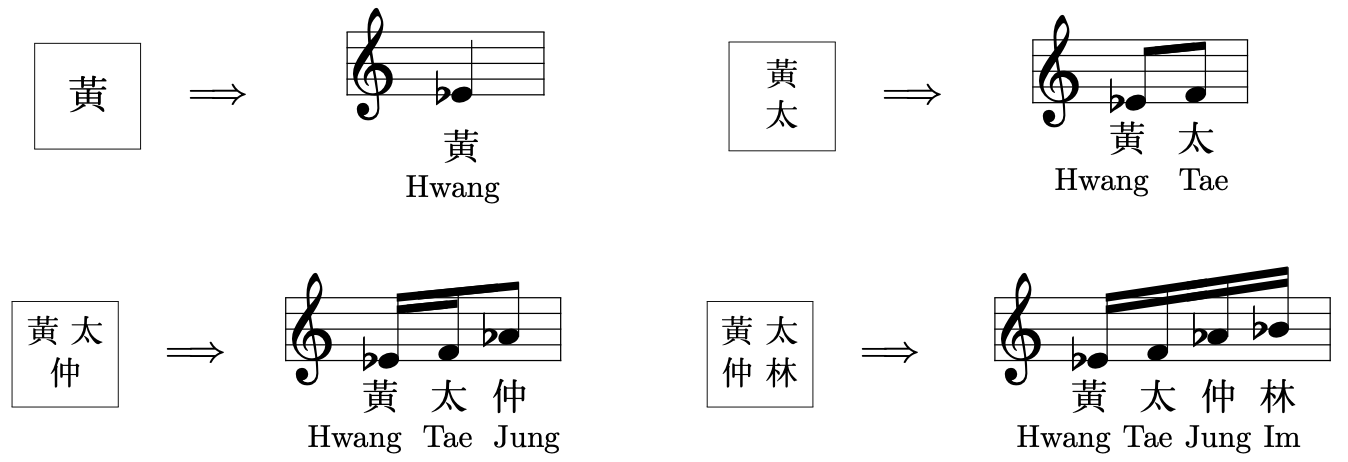}
\caption{Reading pitches in one Jeonggan.}
\label{fig:1jeonggan}
\end{figure}

\begin{figure}[!h]
\centering
\includegraphics[scale=0.6]{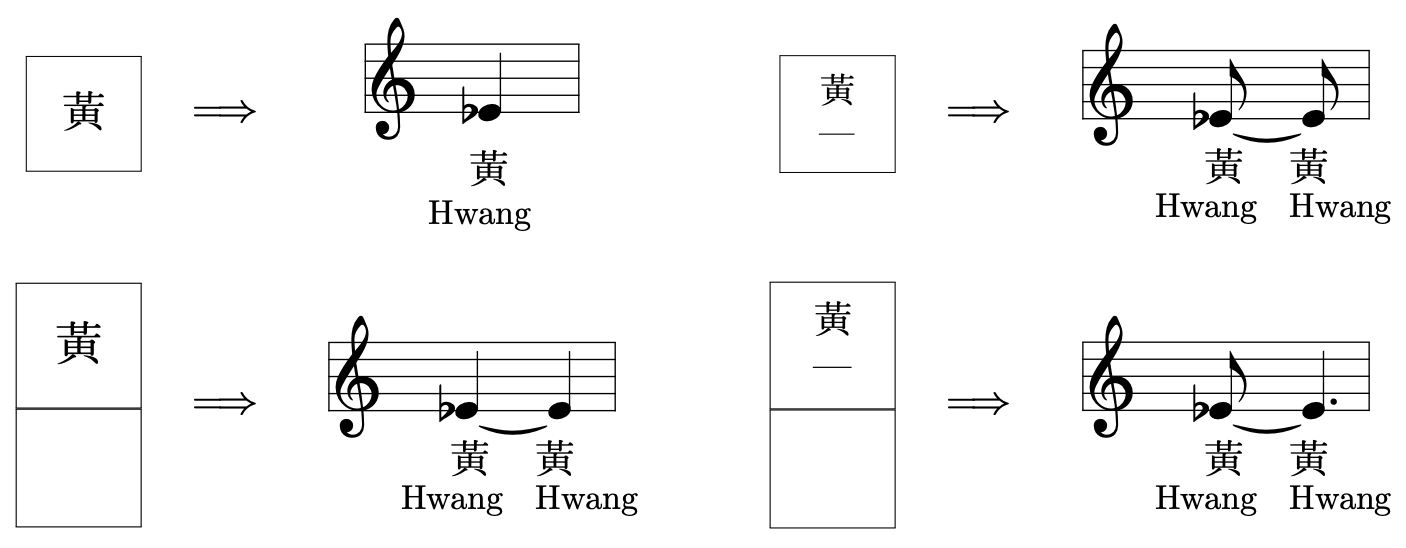}
\caption{Reading the length of a pitch in Jeongganbo.}
\label{fig:pitchlength}
\end{figure}

Jeongganbo was created some time before June of 1447 by Sejong the Great of Joseon dynasty and many musicians in order to overcome the limitations of existing musical notations including Yukbo, Yuljabo and Gongcheokbo \citep{HKY}. It is the first notation method which shows the exact length and height of the sound to be played. In Jeongganbo, each column is usually composed of 32, 16, 12 or 6 squares called Jeonggan (井間). The pitches are displayed by the first letters of each of the 12 Yulmyeong (Hwang, Dae, Tae, Hyeop, Go, Jung, Yu, Im, I, Nam, Mu and Eung) and the length of each note is displayed by the number of Jeonggans. Typically one Jeonggan represents the equivalent of one quarter note \citep{PG}. So, as shown in \figref{fig:1jeonggan} for example, if only one Hwang is placed in one Jeonggan, it counts as one beat; and if Hwang and Tae are both in one Jeonggan {\color{black}{and occupy equal amount of space}}, this divides the rhythm into eighth notes and so on. Within each Jeonggan, the notes are read from left to right and from top to bottom. The rule of reading the length of a pitch in Jeongganbo is illustrated in \figref{fig:pitchlength}. 
%%%%%%%%%%%%%%%%%%%%%%%%%%%%%%
%%%%%%%%%%%%%%%%%%%%%%%%%%%%%%

\begin{figure}[!h]\centering
\includegraphics[scale=0.55]{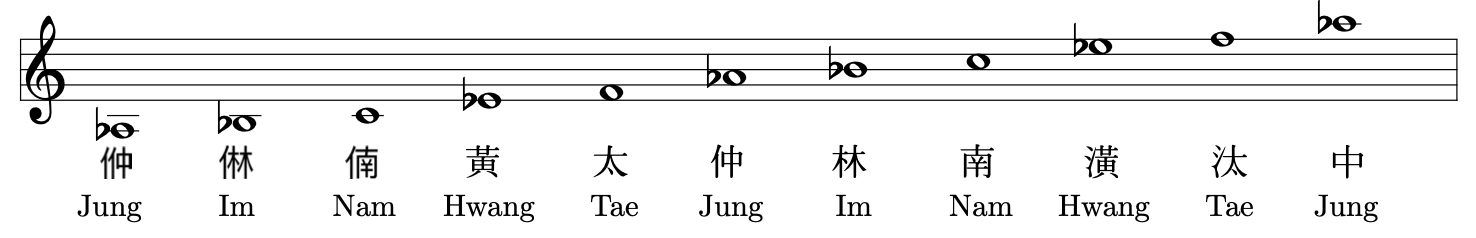}
\caption{Eleven pitches used in Suyeonjang, Songkuyeo and Taryong.}
\label{fig:11pitches}
\end{figure}

 \textcolor{black}{In this paper we study in particular three music pieces Suyeonjang, Songkuyeo and Taryong, mentioned earlier, played by Haegeum instrument. In Suyeonjang and Songkuyeo each column contains 6 Jeonggans, while in Taryong each column contains 12 Jeonggans. In these music pieces five notes including Jung ($G\#3$), Im ($A\#3$), Nam ($C4$), Hwang ($D\#4$) and Tae ($F4$)  are used, where Jung ($G\#3$) can be played two octaves higher and the remaining four notes can be played one octave higher, making a total of eleven pitches, that are $G\#3$, $A\#3$, $C4$, $D\#4$, $F4$, $G\#4$, $A\#4$, $C5$, $D\#5$, $F5$ and $G\#5$, as shown in \figref{fig:11pitches}. The lengths of notes range from 1/6-Jeonggan to 6-Jeonggan.
 }
%\begin{figure}[!h]\centering
%		\startextract
%		\scale{1}
%		\NOTEs\zchar{-10.3}{\hspace{-.13cm}\includegraphics[width=0.36cm]{jung0.png}\hspace{1.cm}\includegraphics[width=0.4cm]{Im0.png}\hspace{1.05cm}\includegraphics[width=0.4cm]{Nam0.png}}\zchar{-9.4}{\hspace{4.05cm}黃\hspace{1.05cm}太\hspace{.9cm}仲\hspace{1.0cm}林}\zchar{-15}{\hspace{-.4cm}{\footnotesize Jung \hspace{.65cm} Im \hspace{.65cm} Nam \hspace{.3cm} Hwang \hspace{.4cm} Tae \hspace{.45cm} Jung \hspace{.6cm} Im}}\wh {_a} \wh {_b} \wh c \wh {_e} \wh f \wh {_h} \wh {_i} \en 
%          	\endextract
%\caption{Seven pitches used in Suyeonjang, \textcolor{red}{Songkuyeo and Taryong.}}
%\label{fig:7pitches}
%\end{figure}

%\begin{figure}[!h]\centering
%		\startextract
%		\scale{1}
%		\NOTEs\zchar{-10.3}{\hspace{-.13cm}\includegraphics[width=0.36cm]{jung0.png}\hspace{1.cm}\includegraphics[width=0.4cm]{Im0.png}\hspace{1.05cm}\includegraphics[width=0.4cm]{Nam0.png}}\zchar{-9.4}{\hspace{4.05cm}黃\hspace{1.05cm}太\hspace{.9cm}仲\hspace{1.0cm}林\hspace{1.05cm}南\hspace{1.0cm}潢 \hspace{.9cm}汰 }\zchar{-15}{\hspace{-.4cm}{\footnotesize Jung \hspace{.65cm} Im \hspace{.65cm} Nam \hspace{.3cm} Hwang \hspace{.4cm} Tae \hspace{.45cm} Jung \hspace{.6cm} Im \hspace{.65cm} Nam\hspace{.4cm} Hwang \hspace{.4cm} Tae }}\wh {_a} \wh {_b} \wh c \wh {_e} \wh f \wh {_h} \wh {_i} \wh {'c} \wh {_e} \wh{f} \en 
%          	\endextract
%\caption{Ten pitches used in Suyeonjang, \textcolor{red}{Songkuyeo and Taryong.}}
%\label{fig:10pitches}
%\end{figure}

%\begin{figure}[!h]\centering
%		\startextract
%		\scale{1}
%		\NOTEs\zchar{-10.3}{\hspace{-.13cm}\includegraphics[width=0.36cm]{jung0.png}\hspace{1.cm}\includegraphics[width=0.4cm]{Im0.png}\hspace{1.05cm}\includegraphics[width=0.4cm]{Nam0.png}}\zchar{-9.4}{\hspace{4.05cm}黃\hspace{1.05cm}太\hspace{.9cm}仲\hspace{1.0cm}林\hspace{1.05cm}南\hspace{1.0cm}潢 \hspace{.9cm}汰 \hspace{.8cm}中 }\zchar{-15}{\hspace{-.4cm}{\footnotesize Jung \hspace{.65cm} Im \hspace{.65cm} Nam \hspace{.3cm} Hwang \hspace{.4cm} Tae \hspace{.45cm} Jung \hspace{.6cm} Im \hspace{.65cm} Nam\hspace{.4cm} Hwang \hspace{.4cm} Tae \hspace{.45cm} Jung }}\wh {_a} \wh {_b} \wh c \wh {_e} \wh f \wh {_h} \wh {_i} \wh {'c} \wh {_e} \wh{f} \wh{_h} \en 
%          	\endextract
%\caption{Eleven pitches used in Suyeonjang, Songkuyeo and Taryong.}
%\label{fig:11pitches}
%\end{figure}

\begin{figure}[!h]
\centering
\includegraphics[scale=0.6]{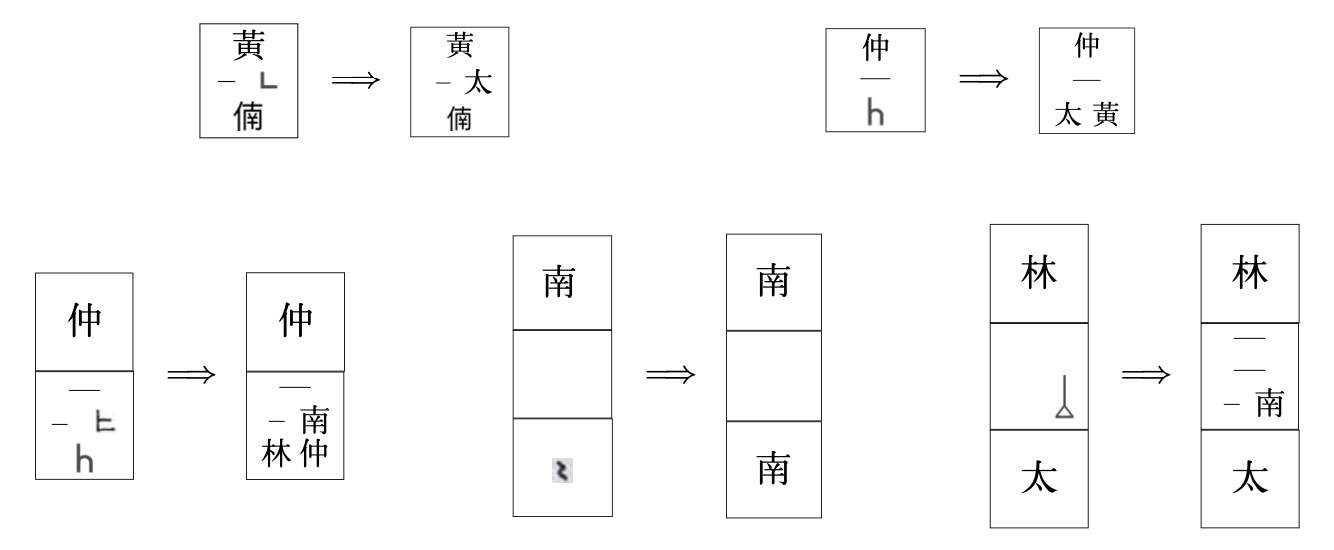}
\caption{Reading musical symbols used in Suyeonjang, Songkuyeo and Taryong.}
\label{fig:abbr}
\end{figure}

 \textcolor{black}{Beside the main letters that show the notes to be played Jeongganbo, in particular Suyeonjang, Songkuyeo and Taryong, various musical symbols are used to notate pitch and length of the note or provide information about how to play the note. For the sake of simplicity but still keeping the main characteristics of the music pieces we have left out \textit{kkumim-eums} (ornamenting tones) and instrument-specific, i.e. Haegeum in our case, notations which give the performer information about special playing techniques such as how to control the bow stick and horsehair. {\color{black}{Development of diverse kkumim-eums is one of the main characteristics of Korean traditional music. Detailed analysis of the effects of kkumim-eums should be made in the future studies}}. Figure \ref{fig:abbr} illustrates the musical symbols used in Suyeonjang, Songkuyeo and Taryong that we interpret. The symbol \includegraphics[height=0.7\fontcharht\font`\B]{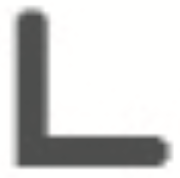} notates one note higher than the previous note, meaning that for example if it appears after note Hwang, then the symbol \includegraphics[height=0.7\fontcharht\font`\B]{an.png} stands for note Tae which is one pitch higher than Hwang. The length of this note is read according to the rule of reading the length of a pitch in Figure \ref{fig:pitchlength}. Symbol \includegraphics[height=\fontcharht\font`\B]{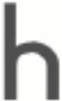} represents two notes, one is one pitch lower than the main note and the other is two pitches lower than the main note. So, if \includegraphics[height=\fontcharht\font`\B]{ah.png}   is placed after note Jung, then two notes Tae and Hwang with equal length should be played. Symbol \includegraphics[height=\fontcharht\font`\B]{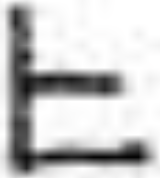} stands for the note that is two pitches higher than the previous note. So if the symbol \includegraphics[height=\fontcharht\font`\B]{at.png}  appears after note Jung then it is supposed to be note Nam. Symbol \includegraphics[height=\fontcharht\font`\B]{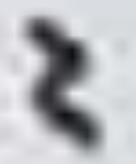} simply stands for a repetition of the previous note. The last symbol \includegraphics[height=\fontcharht\font`\B]{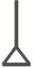} so-called Ingeojil represents a short higher note. So if it appears after note Im then Im is played quite longer followed by a short note Nam before proceeding to the next note.
 }

\section{Construction of music network}
\label{sec:construction}
\textcolor{black}{A conventional approach of studying music is to use the network representation \citep{BW, Gomez, Itz, Liu}}. Let $G = (N, E)$ be the music network that we want to construct, where $N$ is the set of nodes and $E$ is the set of edges in $G$.  We define each node, $n \in N$  in $G$ to be a pair of a pitch and its length as a tuple below 
\begin{equation}
 n = (\mbox{pitch}, \mbox{length}). 
\end{equation}
As mentioned in the previous section, Suyeonjang, \textcolor{black}{Songkuyeo and Taryong consist of eleven pitches including $G\#3$, $A\#3$, $C4$, $D\#4$, $F4$, $G\#4$, $A\#4$, $C5$, $D\#5$, $F5$ and $G\#5$} with the lengths of notes ranging from 1/6-Jeonggan to \textcolor{black}{6}-Jeonggan. For example, if the $i$th node, $n_i$,  is a pair of a pitch, say $G\#3$, and its length, say one Jeonggan, then the node is represented by $ n_i = (G\#3,1)$. Using this construction the network of Suyeonjang, \textcolor{black}{Songkuyeo and Taryong} contains in total $33$, \textcolor{black}{ $37$ and $40$ nodes, respectively}. Two nodes are connected if they occur adjacent in time. Let $e_{ij} \in E$ be the edge  whose end points are $n_i$ and $n_j$. 
The weight or the degree of the edge $e_{ij}$,  {\textcolor{black}{$w_{ij}$}}, between $n_i$ and $n_j$ is the number of occurrences of those two nodes being adjacent in time. 
{\textcolor{black} {For two distinct nodes $n_i$ and $n_j$ with $i<j$, let $p_{ij}$ be the path with the minimum number of edges between $n_i$ and $n_j$ found by using Dijkstra algorithm. 
%The weight on an edge is the number of connections between the end points of the edge.
Then we define the distance $d_{ij}$ between nodes $n_i$ and $n_j$ and form the distance matrix $D = \{d_{ij}\}$ as follows:% \citep{RCG}:
\begin{equation}
	d_{ij} = \left\{ \begin{array}{lc} 
	                {\sum \limits_{e_{kl}\in p_{ij}}w_{kl}^{-1}}, & i < j \\ % \sum _{w(e\in g_{ij})} w^{-1}
	                0, & i = j 
	                \end{array} \right. 
	\quad \text{and} \quad 
	d_{ij} = d_{ji}, i> j,	                
	\label{1}
\end{equation}
where $w_{kl}$ represents the weight of the edge $e_{kl}$ and $p_{ij} = \bigcup e_{kl} $. 
}}

\textcolor{black}{It is clear that if two node $n_k$ and $n_l$ are connected by the edge $e_{kl}$ then $w_{kl} \neq 0$ and further $w_{kl} \ge 1$. Thus $w_{kl}^{-1}$ is well defined in the above distance definition. Also note that since there is no isolated node which is not an end point of any of edges in $E$ (due to the fact that  there is no empty Jeonggan where the music is not played), there always exists at least one path between any two nodes $n_i$ and $n_j$ even if they do not appear adjacently in the whole music, that is, $p_{ij}$ exists, $\forall i<j$. Thus for this definition it is not necessary that $n_i$ and $n_j$ are adjacent. Therefore the definition of the distance Eq. (\ref{1}) is well-defined. This observation is valid at least for those music we consider in this paper, played with Haegeum.   
}
The metric is taken this way because we want to construct a space where the more closely connected nodes in networks are the closer in the space. Therefore, we took the inverse of the weight since we look at networks where larger weights indicate more closely connected relationships between nodes.
%Here note that for this definition it is not necessary that $n_i$ and $n_j$ are adjacent. 
%Further, $g_{ij}$ may not be unique, but the distance $d_{ij}$ should be. We also note that there is no isolated node which is not an end point of any of edges in $E$. This is due to the fact that  there is no empty Jeonggan where the music is not played. It is also obvious that $d_{ij} = d_{ji}$. Thus the definition of the distance Eq. (\ref{1}) is well-defined. This observation is valid at least for those music we consider in this paper, played with Haegeum.  

\begin{remark}
Because of the definition of the distance in Eq. (\ref{1}), we observe that the point cloud obtained from the constructed network of Suyeonjang is in non-Euclidean metric space. Thus, in generating the barcode we use the method \textsf{ExplicitMetricSpace} in Javaplex \citep{ATVJ} which works for a point cloud in an arbitrary, in particular non-Euclidean, metric space. 
\end{remark}

All nodes of Suyeonjang, Songkuyeo and Taryong, according to the node definition above, are provided in Tables \ref{tab:syj_allnodes},  \ref{tab:songkuyeo_allnodes} and  \ref{tab:taryong_allnodes} in Appendix. All nodes  in each table are listed  in ascending order in terms of pitch first then length. 
That is, $n_j$ has a higher pitch than $n_i$  or both have the same pitch but $n_j$ has a longer length than $n_i$ if  $j > i$. 
 Suyeonjang, Songkuyeo and Taryong have, in total,  $33$, $37$ and $40$ nodes, respectively. 

\section{Frequency}  \label{sec:freq}
\begin{figure}[th]
	\centering
		\includegraphics[width=0.3\textwidth]{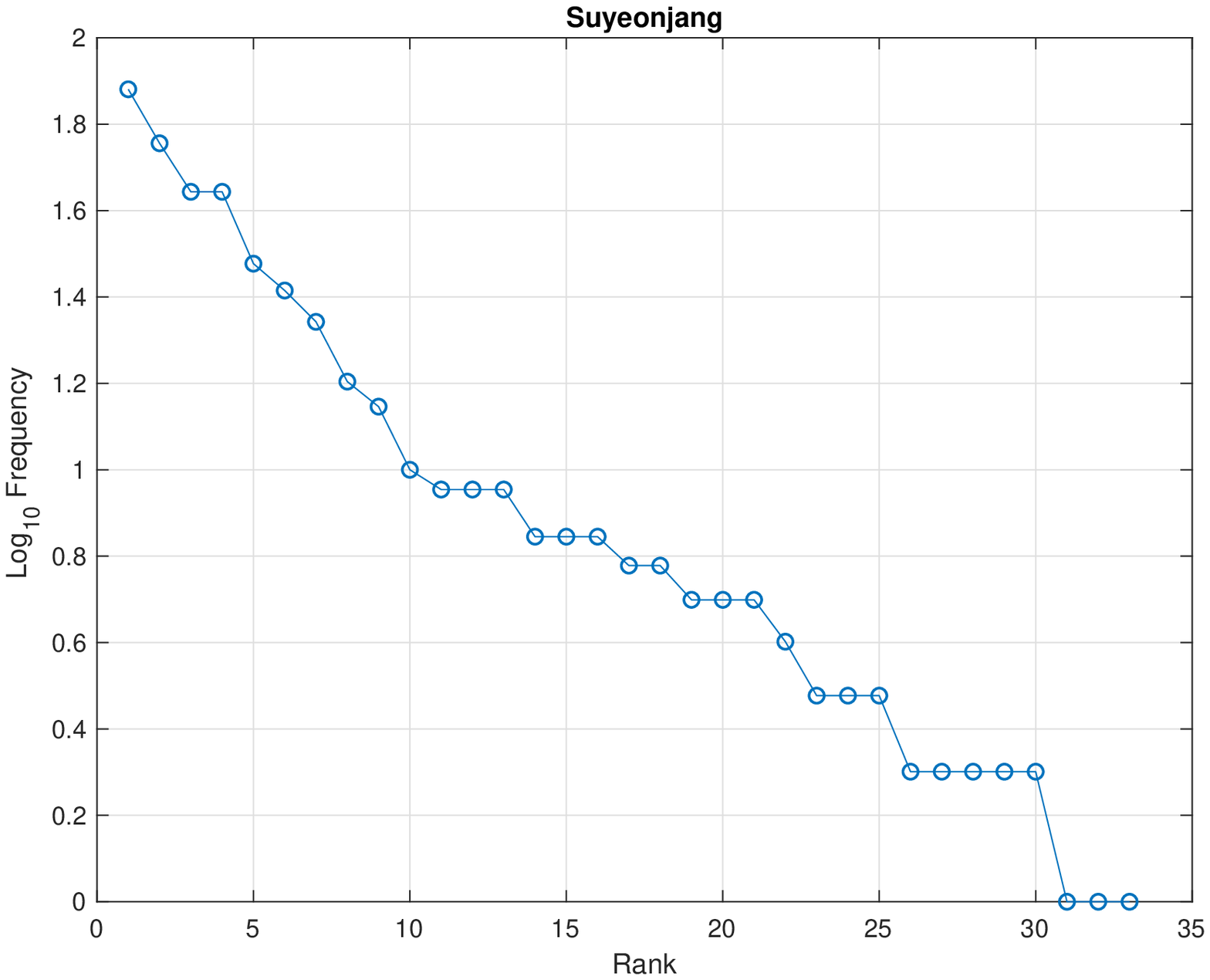}
		\includegraphics[width=0.3\textwidth]{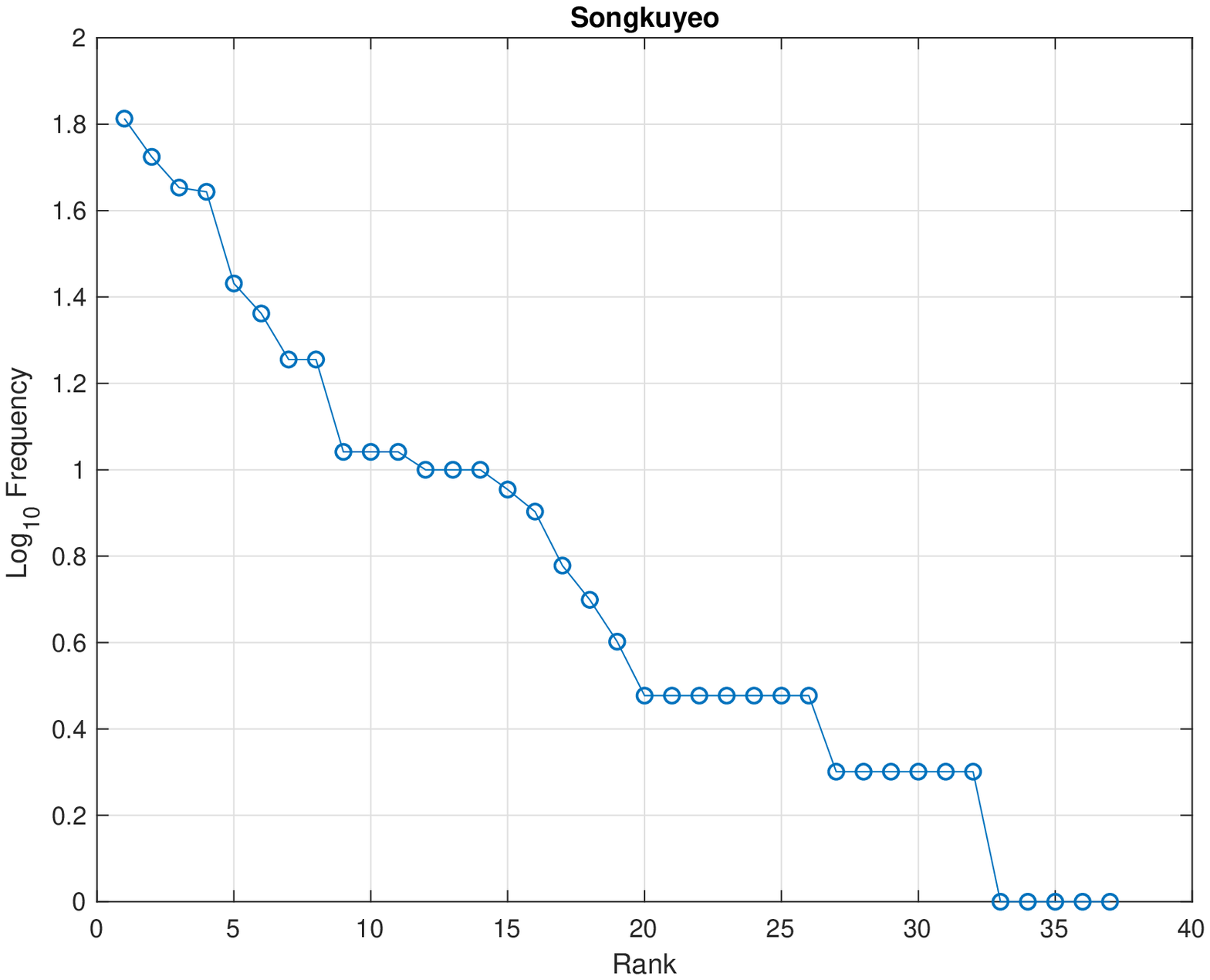}
		\includegraphics[width=0.3\textwidth]{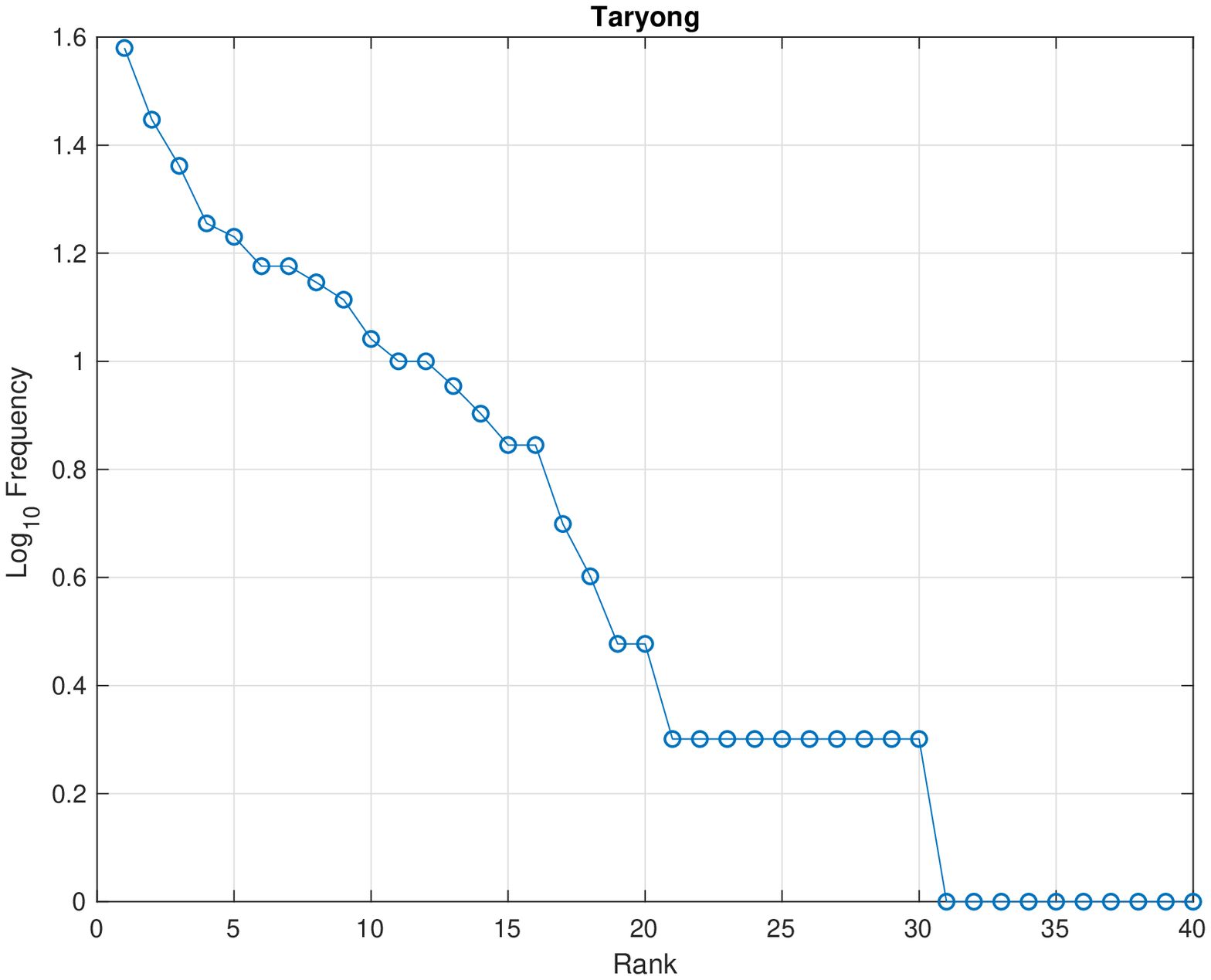}
	     \caption{Frequency versus node. The horizontal axis represents the rank and the vertical axis the frequency in logarithmic scale. Left: Suyeonjang. Middle: Songkuyeo. Right: Taryong. Note that the minimum frequency is unity as the minimum occurrence of a particular node is at least one by definition. }
		%$\frac{dE}{dt}$ for $\eta = 0.9805$. $\frac{dE}{dt}$ is increasing strictly. (fig:syj_numofcyc805)}
	 \label{figure1}
\end{figure}

We first provide the frequency of the occurrence of each node in music. Table \ref{tab:table1} shows the node frequency for Suyeonjang, Songkuyeo and Taryong, left to right, respectively. For each music, the left column shows the node number and the right column the corresponding occurrence frequency. The frequencies are listed from first to last in rank in the table. 
Figure \ref{figure1}  shows the frequency versus rank in semi-logarithmic scale. The horizontal axis represents the rank and the vertical axis the frequency in logarithmic scale. Note that the frequency is a positive integer by definition. As shown in the figure, the frequency decays more exponentially than algebraically. 

\begin{table}[]
  \begin{center}
    \caption{Frequency versus node.}
    \label{tab:table1}
    {\footnotesize{
    \begin{tabular}{| c | |c c|| c c|| c c|} % <-- Alignments: 1st column left, 2nd middle and 3rd right, with vertical lines in between
          \hline
    {rank}&   {Suyeonjang} &  & {Songkuyeo} &  & {Taryong} &  \\
 %     $\alpha$ & $\beta$ & $\gamma$ \\
      \hline
1 & $n_{18}$ & 76 & $n_{20}$ & 65 & $n_{16}$ & 38 \\ 
%2 & $n_{6}$ & 57 & $n_{32}$ & 53 & $n_{11}$ & 28 \\ 
2 & $n_{6}$ & 57 & $n_{31}$ & 53 & $n_{11}$ & 28 \\ 
3 & $n_{11}$ & 44 & $n_{13}$ & 45 & $n_{13}$ & 23 \\ 
%4 & $n_{22}$ & 44 & $n_{26}$ & 44 & $n_{27}$ & 18 \\ 
4 & $n_{22}$ & 44 & $n_{26}$ & 44 & $n_{26}$ & 18 \\ 
%5 & $n_{1}$ & 30 & $n_{8}$ & 27 & $n_{30}$ & 17 \\ 
5 & $n_{1}$ & 30 & $n_{8}$ & 27 & $n_{29}$ & 17 \\ 
%6 & $n_{20}$ & 26 & $n_{18}$ & 23 & $n_{21}$ & 15 \\ 
6 & $n_{20}$ & 26 & $n_{18}$ & 23 & $n_{31}$ & 15 \\ 
7 & $n_{27}$ & 22 & $n_{4}$ & 18 & $n_{28}$ & 15 \\ 
%8 & $n_{3}$ & 16 & $n_{34}$ & 18 & $n_{3}$ & 14 \\ 
8 & $n_{3}$ & 16 & $n_{33}$ & 18 & $n_{3}$ & 14 \\ 
9 & $n_{28}$ & 14 & $n_{6}$ & 11 & $n_{18}$ & 13 \\ 
10 & $n_{12}$ & 10 & $n_{16}$ & 11 & $n_{15}$ & 11 \\ 
11 & $n_{16}$ & 9 & $n_{25}$ & 11 & $n_{12}$ & 10 \\ 
%12 & $n_{26}$ & 9 & $n_{19}$ & 10 & $n_{23}$ & 10 \\ 
12 & $n_{26}$ & 9 & $n_{19}$ & 10 & $n_{22}$ & 10 \\ 
13 & $n_{31}$ & 9 & $n_{24}$ & 10 & $n_{6}$ & 9 \\ 
%14 & $n_{2}$ & 7 & $n_{27}$ & 10 & $n_{33}$ & 8 \\ 
14 & $n_{2}$ & 7 & $n_{27}$ & 10 & $n_{32}$ & 8 \\ 
15 & $n_{4}$ & 7 & $n_{28}$ & 9 & $n_{17}$ & 7 \\ 
%16 & $n_{23}$ & 7 & $n_{33}$ & 8 & $n_{20}$ & 7 \\ 
16 & $n_{23}$ & 7 & $n_{32}$ & 8 & $n_{20}$ & 7 \\ 
17 & $n_{9}$ & 6 & $n_{2}$ & 6 & $n_{4}$ & 5 \\ 
18 & $n_{10}$ & 6 & $n_{15}$ & 5 & $n_{9}$ & 4 \\ 
19 & $n_{5}$ & 5 & $n_{7}$ & 4 & $n_{0}$ & 3 \\ 
20 & $n_{8}$ & 5 & $n_{11}$ & 3 & $n_{14}$ & 3 \\ 
21 & $n_{13}$ & 5 & $n_{12}$ & 3 & $n_{2}$ & 2 \\ 
22 & $n_{0}$ & 4 & $n_{14}$ & 3 & $n_{5}$ & 2 \\ 
23 & $n_{7}$ & 3 & $n_{17}$ & 3 & $n_{7}$ & 2 \\ 
24 & $n_{17}$ & 3 & $n_{21}$ & 3 & $n_{8}$ & 2 \\ 
25 & $n_{19}$ & 3 & $n_{23}$ & 3 & $n_{19}$ & 2 \\ 
%26 & $n_{21}$ & 2 & $n_{36}$ & 3 & $n_{22}$ & 2 \\ 
26 & $n_{21}$ & 2 & $n_{35}$ & 3 & $n_{21}$ & 2 \\ 
%27 & $n_{25}$ & 2 & $n_{0}$ & 2 & $n_{29}$ & 2 \\ 
27 & $n_{25}$ & 2 & $n_{0}$ & 2 & $n_{27}$ & 2 \\ 
%28 & $n_{29}$ & 2 & $n_{3}$ & 2 & $n_{34}$ & 2 \\ 
28 & $n_{29}$ & 2 & $n_{3}$ & 2 & $n_{33}$ & 2 \\ 
%29 & $n_{30}$ & 2 & $n_{9}$ & 2 & $n_{35}$ & 2 \\ 
29 & $n_{30}$ & 2 & $n_{9}$ & 2 & $n_{34}$ & 2 \\ 
%30 & $n_{32}$ & 2 & $n_{10}$ & 2 & $n_{36}$ & 2 \\ 
30 & $n_{32}$ & 2 & $n_{10}$ & 2 & $n_{35}$ & 2 \\ 
%31 & $n_{14}$ & 1 & $n_{31}$ & 2 & $n_{1}$ & 1 \\ 
31 & $n_{14}$ & 1 & $n_{36}$ & 2 & $n_{1}$ & 1 \\ 
%32 & $n_{15}$ & 1 & $n_{35}$ & 2 & $n_{10}$ & 1 \\ 
32 & $n_{15}$ & 1 & $n_{34}$ & 2 & $n_{10}$ & 1 \\ 
%33 & $n_{24}$ & 1 & $n_{1}$ & 1 & $n_{24}$ & 1 \\ 
33 & $n_{24}$ & 1 & $n_{1}$ & 1 & $n_{23}$ & 1 \\ 
%34 & & & $n_{5}$ & 1 & $n_{25}$ & 1 \\ 
34 & & & $n_{5}$ & 1 & $n_{24}$ & 1 \\ 
%35 & & & $n_{22}$ & 1 & $n_{26}$ & 1 \\
35 & & & $n_{22}$ & 1 & $n_{25}$ & 1 \\
%36 & & & $n_{29}$ & 1 & $n_{31}$ & 1 \\
36 & & & $n_{29}$ & 1 & $n_{30}$ & 1 \\
%37 & & & $n_{30}$ & 1 & $n_{32}$ & 1 \\
37 & & & $n_{30}$ & 1 & $n_{38}$ & 1 \\
%38 & & &  & & $n_{37}$ & 1 \\ 
38 & & &  & & $n_{36}$ & 1 \\ 
%39 & & &  & & $n_{38}$ & 1 \\
39 & & &  & & $n_{37}$ & 1 \\
40 & & &  & & $n_{39}$ & 1 \\ 
 \hline
       \end{tabular}
       }}
  \end{center}
\end{table}

\section{Analysis of the barcodes} \label{sec:analysis}
In this section we analyze the barcode and give an interpretation of the persistence intervals in the first dimension for Suyeonjang, Songkuyeo and Taryong. 

\subsection{Suyeonjang}
\begin{figure}[h]
    \begin{center}
    \includegraphics[width=0.5\textwidth]{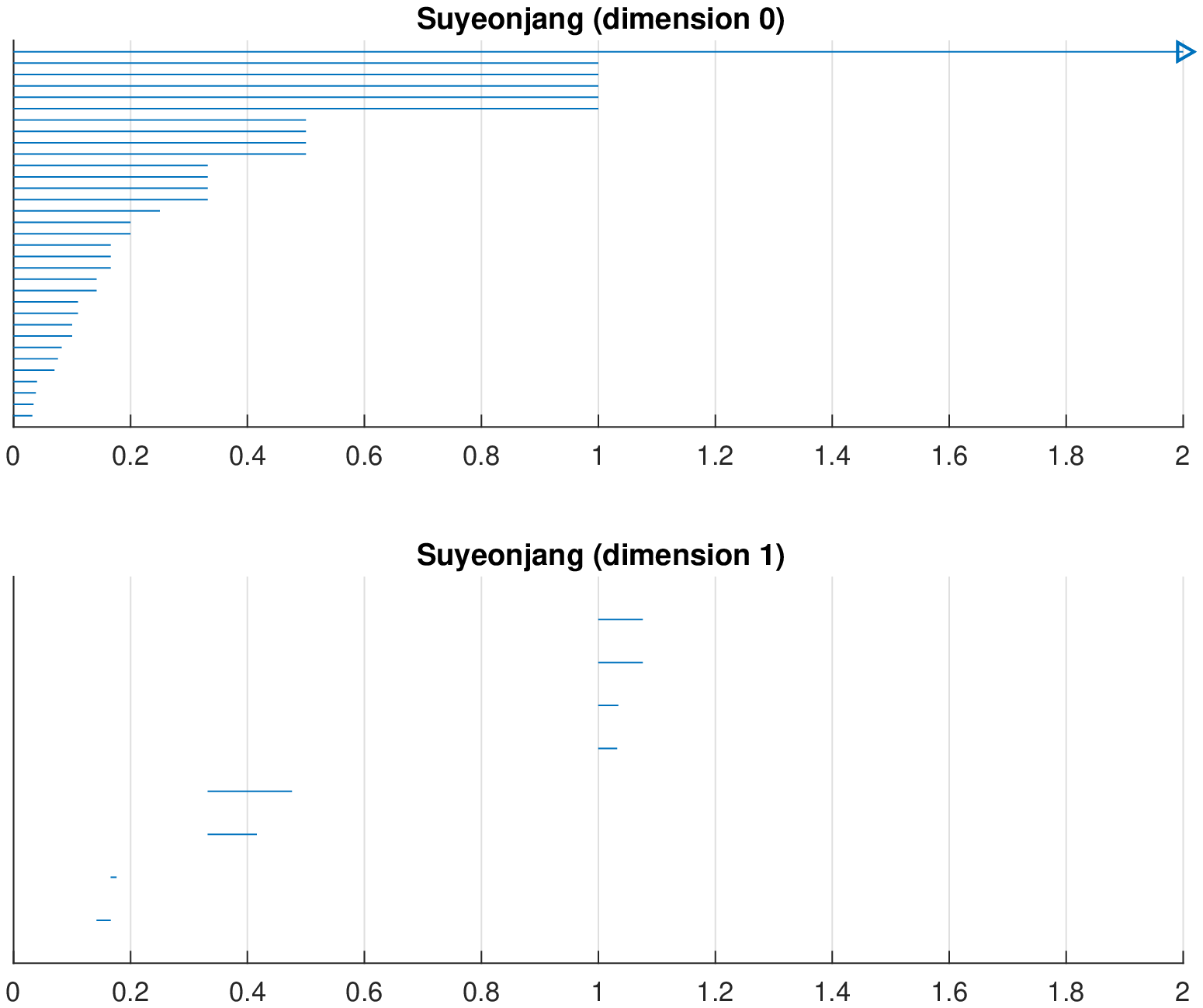}
    \caption{Barcode of Suyeonjang using Vietoris-Rips method. Top: 0-D barcode. Bottom: 1-D barcode. In 1D barcode, we observe 8 non-zero persistence implying 8 cycles in $G$.}
    \label{fig:syj_barcode}
    \end{center}
\end{figure}

The barcode for Suyeonjang generated by Javaplex with Vietoris-Rips method is given in \figref{fig:syj_barcode}. 
% syj barcode
In \figref{fig:syj_barcode} the horizontal axis is the filtration value $\tau$. Vertically we have multiple intervals that correspond to generators of the homology groups. In the zeroth dimension we have $33$ generators that correspond to $33$ components when $\tau$ is small, which eventually are connected into a single component when $\tau=1$. The $33$ components are actually those $33$ nodes defined in Suyeonjang. All these components constitute a single component because of the fact that any node in the network connects at least one time with another node, which means that at most when distance $d=1$ all nodes in the network are connected. In the first dimension we see $8$ generators which topologically correspond to eight cycles. 
A natural question is whether these cycles are related to the repetition of music melodies. In the following we will go into detailed analysis about the musical meaning of these cycles.

The persistence algorithm computing intervals is used to find a representative cycle for each interval. The annotated intervals computed by the method \textsf{computeAnnotatedIntervals} in Javaplex tell us what the nodes in the intervals of persistence are. In the one dimension case, the annotated intervals consist of the components in the loops generated in the process of filtration. 

\figref{fig:syj_cycles} shows $8$ cycles identified by TDA corresponding to $8$ persistence intervals in dimension one of the barcode. The order of cycles is the order of their appearance in the barcode. 
%That is, the earlier value of $\tau$ the 1D barcode has the lower number is assigned to the corresponding cycle. For example, the birth of Cycle $i$ is earlier than the birth of Cycle $j$ if $i < j$.
That is, the earlier \textcolor{black}{ the 1D barcode dies} the lower number is assigned to the corresponding cycle.
For example, the \textcolor{black}{death} of Cycle $i$ is earlier than the \textcolor{black}{death} of Cycle $j$ if $i < j$. Note that this order is different from the order of their appearance in the actual music.  In the figure, each cycle is shown with {\textcolor{black}{the persistence interval,}} the node information (its node number and note), edge weight ({\textcolor{black}{in normal size in blue), distance between node (in small size in blue in brackets)}} and the average weight (in red in center) which is the simple mean of all edge weights. As shown in the figure, the minimum number of nodes that constitute a cycle is $4$ and the maximum number is $6$. The average node number is $4.625$. The average weight is $9.39375$. Cycle 2 {\color{black}{corresponds}} to the shortest persistence in 1D barcode. Cycle 4 {\color{black}{corresponds}} to the longest persistence in 1D barcode.
 %Cycles $6, 7$ and $8$ do not appear in actual music in their whole consecutive form. They are corresponding to those persistence starting from $d = 1$ in 1D barcode.  
 The information of corresponding music notes is given in \tabref{tab:syj_nodeinfo}. {\color{black}{It should be kept}} in mind that Suyeonjang has in total $33$ nodes, and the cycles show which node combinations are of interest.
 
%%%%%%%%%%%%%%%%%%%%%%%%%%%%%
\include{syj_cycles}
\include{syj_nodeinfo}
%%%%%%%%%%%%%%%%%%%%%%
%%%%%%%%%%%%%%%%%%%%%%

The first attempt is to find in the music sheet the sequence of music notes appearing in each cycle. We have found that except for cycles 6, 7 and 8, five remaining cycles indeed represent consecutive sequences of notes in Suyeonjang. Among them two sequences of music notes actually make up two {\textcolor{black}{closed}} cycles 4 and 5. They appear in Suyeonjang in the following orders 
%sequence 4.
            \begin{center}\begin{tikzpicture}[scale = 1]
                \tikzstyle{every node}=[draw,shape=circle,inner sep=0pt,minimum size=.6cm,inner sep=0pt,minimum size=.6cm];
                \node[fill=blue!99] (v1) at (1,0) {\textcolor{white}{\includegraphics[width=0.36cm]{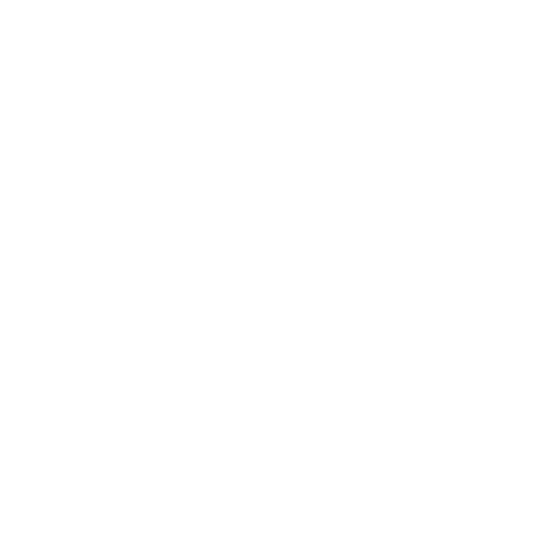}}};%{$6$};
                \node[fill=blue!99] (v2) at (2,0) {\textcolor{white}{\includegraphics[width=0.36cm]{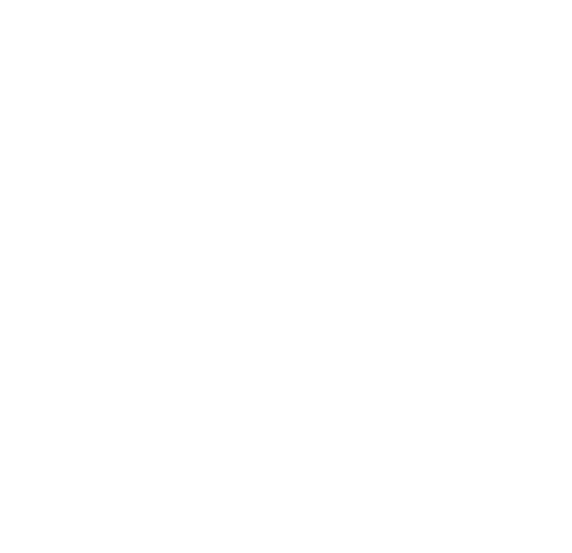}}};%{$11$};
                \node[fill=red!30] (v3) at (3,0) {\textcolor{black}{\includegraphics[width=0.36cm]{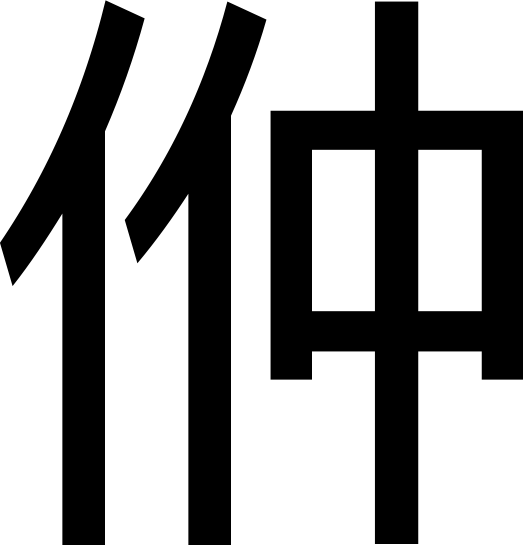}}};%{$2$};
                \node[fill=orange!99] (v4) at (4,0) {\textcolor{white}{\includegraphics[width=0.36cm]{Imw.png}}};%{$7$};
                \node[fill=teal!60] (v5) at (5,0) {\textcolor{black}{\includegraphics[width=0.36cm]{jung0.png}}};%{$0$};
                \node[fill=blue!99] (v6) at (6,0) {\textcolor{white}{\includegraphics[width=0.36cm]{Imw.png}}};%{$6$};
                \node[draw=none] (v01) at (1.1,-0.5) {$n_{6}$};
                \node[draw=none] (v02) at (2.1,-0.5) {$n_{11}$};
                \node[draw=none] (v03) at (3.1,-0.5) {$n_{2}$};
                \node[draw=none] (v04) at (4.1,-0.5) {$n_{7}$};
                \node[draw=none] (v05) at (5.1,-0.5) {$n_{0}$};
                \node[draw=none] (v06) at (6.1,-0.5) {$n_{6}$};
                \draw (v1) -- (v2)
                    (v2) -- (v3)
                    (v3) -- (v4)
                    (v4) -- (v5)
                    (v5) -- (v6);
            \end{tikzpicture}\end{center}  
for cycle 4 (appearing three times) and
%sequence 5.
            \begin{center}\begin{tikzpicture}[scale = 1]
                \tikzstyle{every node}=[draw,shape=circle,inner sep=0pt,minimum size=.6cm,inner sep=0pt,minimum size=.6cm];
                \node[fill=blue!99] (v1) at (1,0) {\textcolor{white}{\includegraphics[width=0.36cm]{Imw.png}}};%{$6$};
                \node[fill=blue!99] (v2) at (2,0) {\textcolor{white}{黃}};%{$18$};
                \node[fill=blue!30] (v3) at (3,0) {\textcolor{black}{太}};%{$21$};
                \node[fill=teal!60] (v4) at (4,0) {\textcolor{black}{黃}};%{$16$};
                \node[fill=blue!99] (v5) at (5,0) {\textcolor{white}{\includegraphics[width=0.36cm]{Imw.png}}};%{$6$};
                \node[fill=blue!99] (v6) at (6,0) {\textcolor{white}{黃}};%{$18$};
                \node[draw=none] (v01) at (1.1,-0.5) {$n_{6}$};
                \node[draw=none] (v02) at (2.1,-0.5) {$n_{18}$};
                \node[draw=none] (v03) at (3.1,-0.5) {$n_{21}$};
                \node[draw=none] (v04) at (4.1,-0.5) {$n_{16}$};
                \node[draw=none] (v05) at (5.1,-0.5) {$n_{6}$};
                \node[draw=none] (v06) at (6.1,-0.5) {$n_{18}$};
                \draw (v1) -- (v2)
                    (v2) -- (v3)
                    (v3) -- (v4)
                    (v4) -- (v5)
                    (v5) -- (v6);
            \end{tikzpicture}\end{center} 
for cycle 5 (appearing one time), respectively. Note sequences in cycles 2 and 3 are also found in the music in the following orders
%sequence 2.
            \begin{center}\begin{tikzpicture}[scale = 1]
                \tikzstyle{every node}=[draw,shape=circle,inner sep=0pt,minimum size=.6cm,inner sep=0pt,minimum size=.6cm];
                \node[fill=blue!99] (v1) at (1,0) {\textcolor{white}{\includegraphics[width=0.36cm]{Imw.png}}};%{$6$};
                \node[fill=orange!99] (v2) at (2,0) {\textcolor{white}{\includegraphics[width=0.36cm]{namw.png}}};%{$12$};
                \node[fill=teal!60] (v3) at (3,0) {\textcolor{black}{\includegraphics[width=0.36cm]{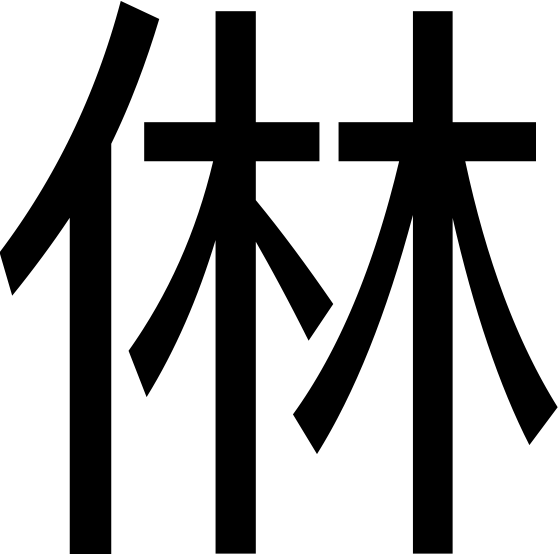}}};%{$3$};
                \node[fill=blue!99] (v4) at (4,0) {\textcolor{white}{黃}};%{$18$};
                \node[draw=none] (v01) at (1.1,-0.5) {$n_{6}$};
                \node[draw=none] (v02) at (2.1,-0.5) {$n_{12}$};
                \node[draw=none] (v03) at (3.1,-0.5) {$n_{3}$};
                \node[draw=none] (v04) at (4.1,-0.5) {$n_{18}$};
                \draw (v1) -- (v2)
                    (v2) -- (v3)
                    (v3) -- (v4);
            \end{tikzpicture}\end{center}             
for cycle 2 (appearing two times) and 
%sequence 3.
            \begin{center}\begin{tikzpicture}[scale = 1]
                \tikzstyle{every node}=[draw,shape=circle,inner sep=0pt,minimum size=.6cm,inner sep=0pt,minimum size=.6cm];
                \node[fill=teal!60] (v1) at (1,0) {\textcolor{black}{太}};%{$20$};
                \node[fill=blue!30] (v2) at (2,0) {\textcolor{black}{仲}};%{$26$};
                \node[fill=orange!99] (v3) at (3,0) {\textcolor{white}{太}};%{$23$};
                \node[fill=teal!60] (v4) at (4,0) {\textcolor{black}{黃}};%{$16$};
                \node[fill=blue!99] (v5) at (5,0) {\textcolor{white}{太}};%{$22$};
                \node[draw=none] (v01) at (1.1,-0.5) {$n_{20}$};
                \node[draw=none] (v02) at (2.1,-0.5) {$n_{26}$};
                \node[draw=none] (v03) at (3.1,-0.5) {$n_{23}$};
                \node[draw=none] (v04) at (4.1,-0.5) {$n_{16}$};
                \node[draw=none] (v05) at (5.1,-0.5) {$n_{22}$};
                \draw (v1) -- (v2)
                    (v2) -- (v3)
                    (v3) -- (v4)
                    (v4) -- (v5);
            \end{tikzpicture}\end{center} 
for cycle 3 (appearing one time), respectively. Although these two sequences do not make up closed cycles, they do preserve the order of edge connectivity of the cycles. For cycle~1 we found two sequences of the same notes but in a slightly different order
%sequence 1.
            \begin{center}\begin{tikzpicture}[scale = 1]
                \tikzstyle{every node}=[draw,shape=circle,inner sep=0pt,minimum size=.6cm,inner sep=0pt,minimum size=.6cm];
                \node[fill=teal!60] (v1) at (1,0) {\textcolor{black}{太}};%{$20$};
                \node[fill=blue!99] (v2) at (2,0) {\textcolor{white}{仲}};%{$27$};
                \node[fill=blue!99] (v3) at (3,0) {\textcolor{white}{黃}};%{$18$};
                \node[fill=blue!99] (v4) at (4,0) {\textcolor{white}{太}};%{$22$};
                \node[draw=none] (v01) at (1.1,-0.5) {$n_{20}$};
                \node[draw=none] (v02) at (2.1,-0.5) {$n_{27}$};
                \node[draw=none] (v03) at (3.1,-0.5) {$n_{18}$};
                \node[draw=none] (v04) at (4.1,-0.5) {$n_{22}$};
                \draw (v1) -- (v2)
                    (v2) -- (v3)
                    (v3) -- (v4);
            \end{tikzpicture}\end{center}

In summary, by using TDA tools we obtain a set of $8$ persistence intervals that can be interpreted as $8$ cycles in dimension one. Our approach is to find sequences of notes in the music that correspond to these cycles. We have found that out of the $8$ cycles five of them appear as consecutive sequences of notes in Suyeonjang (single or multiple times) and the other three are not found. The order in which notes appear in the music is not necessarily the same as the order of edge connectivity of the cycles. %{\textcolor{black}{By this approach we do not have a full explanation of the meaning of the cycles in term of music yet. However, we believe that they contain information about the important melodies of the music and can be used later in other tasks such as generating music using machine learning algorithms. In the next section we discuss our next approach that is expected to give more information about persistent intervals.}}
%%%%%%%%%%%%
%%%%%%%%%%%%

\begin{figure}[!h]
	\begin{center}
            	\includegraphics[height = 5cm, width = \textwidth]{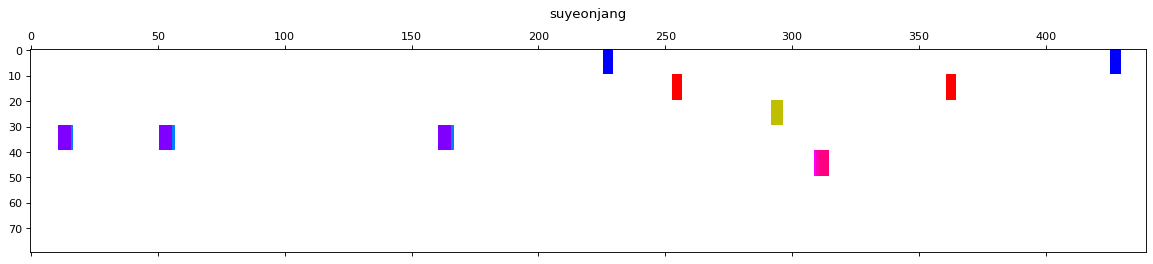}
		\caption{Consecutive sequences of notes corresponding to Cycles 1, 2, 3, 4, 5 found in Suyeonjang. The horizontal axis is the time sequence of music flow and the vertical axis is for the cycles. The top bars represents Cycle 1 and the bottom Cycle 5. The width of each bar is the number of nodes contained in each cycle. The height of each bar is only for the visual distinction purpose. Note that Cycles 6, 7 and 8 are not appearing.  } 
			\label{fig:7}
	\end{center}
\end{figure}

\figref{fig:7} shows the $5$ cycles appearing in Suyeonjang with time. The horizontal axis represents the time sequence the music flows and the vertical axis represents the cycle number, from $1$ to $5$. {\textcolor{black}{ Ignore the number displayed in the axis. }} The width of each is the total number of nodes involved in the cycle. The height of each bar in the figure has no meaning but is displayed only for the visual distinction purpose. {\textcolor{black}{The monotonicity means that the sequence is the same as the cycle sequence. The variance in color means the slight variation of the cycle sequence when appearing.}} For example, the figure shows that Cycle 4 (the $4$th bar from top) appears three times in the first half of music while the rest cycles appear in the second half of the music. We observe that the occurrence of the cycles in the complete form is sparse. 
% Suyeonjang

Let $s$ be a positive integer. Consider a music piece $\mathcal{O}$ composed of $d$ notes that flows in the following order $\mathcal{L} = \{ \nu_1, \ldots, \nu_d\}$ and assume that the barcode for it in the first dimension consists of $k$ generators which topologically correspond to $k$ Cycles, $C_1, \ldots, C_k$. We define the Overlap matrix on $s$-scale for $\mathcal{O}$ as follows.
\definition{Matrix $M_{k\times d}^s=\{m_{ij}^s\}$ is called the Overlap matrix on $s$-scale for $\mathcal{O}$ if it satisfies the following conditions 
%%%%%%%
\[
m_{ij}^s = \begin{cases}
	1, & \text{ if } \exists~ t,l \geq 0 \text{ satisfying } t + l \geq s-1 \text{ such that }\\ 	    
	& \hspace{0.5cm}   \nu_{j-l}, \nu_{j-l-1}, \ldots, \nu_j, \ldots, \nu_{j+t-1}, \nu_{j+t} \in C_i,\\
	0, & \text{ otherwise,}
	\end{cases}
\]
for all $i = 1,\ldots, k$; $j = 1,\ldots,d$.
}\label{def:overlapmat}

In other words, a binary matrix  $M_{k\times d}^s=\{m_{ij}^s\}$, i.e., a matrix whose entries are either $0$ or $1$, belongs to $s$-scale if and only if on each row of $M_{k\times d}^s$ any entry equal to $1$ should be staying in a consecutive sequence of length at least $s$ columns that equal to $1$.

\begin{figure}[!h]
	\begin{center}
            	\includegraphics[height = 4cm, width = \textwidth]{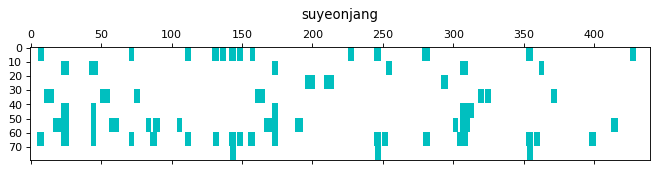}
		\caption{Overlap matrix on $4$-scale for Suyeonjang. This shows consecutive sequences of at least $4$ nodes involved in cycles of Suyeonjang. Similar plot as \figref{fig:7}. Those cycles are displayed when at least $4$ nodes in the cycle are appearing in the music.  Notice that Cycles $6, 7$ and $8$ are appearing in the figure.  } 
		\label{fig:syj_min4notes}
	\end{center}
\end{figure}

\figref{fig:syj_min4notes} shows the Overlap matrix on $4$-scale for Suyeonjang. This plot is similar to  \figref{fig:7},
%\figref{fig:syj_min4notes} shows the similar plot as \figref{fig:7}, 
the cycle occurrence with respect to music flow. The only difference is that each cycle is displayed in the figure whenever at least $4$ nodes in the cycle are appearing in the music. The order may not be the same for each occurrence. 
%The spectrum shown in the width shows the variation of the order of nodes in the cycle when appearing in the music. 
We observe that the occurrence is now rather dense compared to \figref{fig:7}. We also observe that the invisible Cycles $6, 7$ and $8$ in \figref{fig:7} are now appearing in the figure. In fact, the number of the occurrences of Cycle 6 is not small. Also notice that multiple cycles can be found simultaneously. 
%%%

\begin{figure}[!h]
	\begin{center}
	\includegraphics[width=\textwidth]{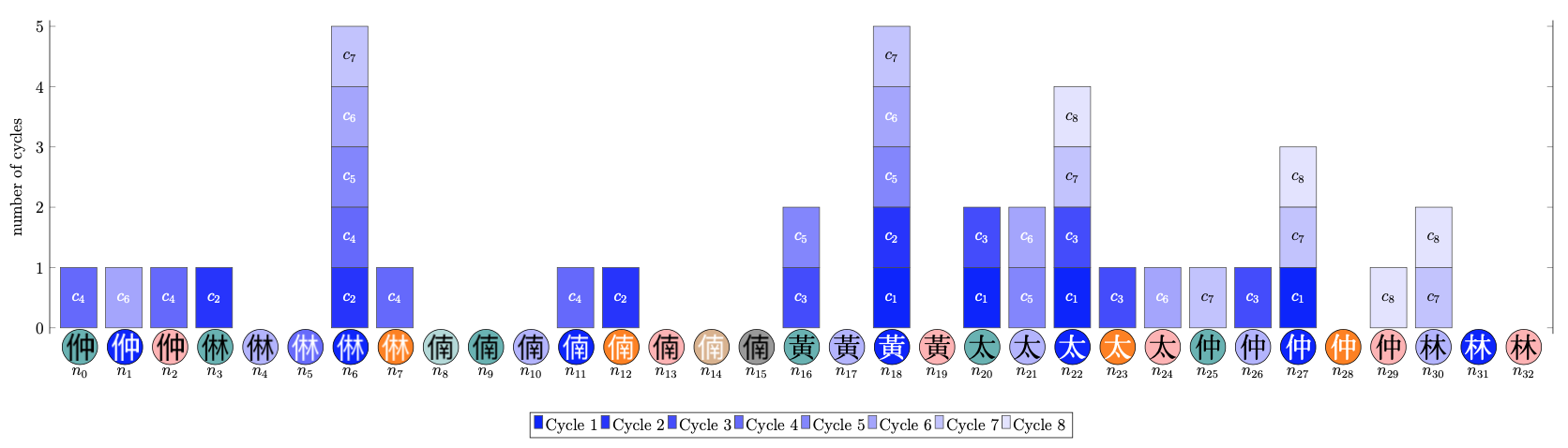}
		\caption{Number of cycles that each node in Suyeonjang belongs to. The horizontal axis shows all nodes in Suyeonjang and the vertical axis shows the total number of cycles that each node belongs to and the actual cycle number.  } 
				\label{fig:syj_numofcyc}
	\end{center}
\end{figure}

\figref{fig:syj_numofcyc} shows how many cycles are there that each of $33$ nodes in Suyeonjang in  \tabref{tab:syj_allnodes} belongs to. The horizontal axis shows all $33$ nodes in ascending order in terms of pitch and the vertical axis shows the total number of cycles that the node belongs to and cycle number together. For examples, each of the first three nodes $n_0, n_1, n_2, n_3$ belongs to only one cycle, i.e. Cycle 4 (c4), Cycle 6 (c6), Cycle 4 (c4) and Cycle 2 (c2) while $n_5$ and $n_6$ do not belong to any of cycles and $n_7$ belongs to $5$ cycles, Cycle 2 (c2), Cycle 4 (c4), Cycle 5 (c5), Cycle 6 (c6) and Cycle 7 (c7). 
%Note that the node number $m_j$ is the node number when all nodes from Suyeonjang, Songkuyeo and Taryong are merged and listed in ascending order in terms of pitch. For example, the node $n_5$ in Suyeonjang becomes the node $m_7$ in the set of all nodes from Suyeonjang, Songkuyeo and Taryong. 

%
% And \figref{fig:10} shows the actual cycle numbers that the node belongs to. For example, the nodes $n_0, n_1, n_2$ and  $n_3$ belong to Cycle 4, Cycle 6, Cycle 4 and Cycle 2, respectively while $n_6$ belongs to Cycle 2, Cycle 4, Cycle 5, Cycle 6 and Cycle 7.  
%%%

%\begin{figure}[!h]
%	\begin{center}
%            	\includegraphics[height = 7cm, width = .7\textwidth]{syj_cyclecount1_pit.png}
%		\caption{Number of cycles that each node in Suyeonjang belongs to. } 
%				\label{fig:syj_numofcyc}
%	\end{center}
%\end{figure}
%%%%
%\begin{figure}[!h]
%	\begin{center}
%            	\includegraphics[height = 7cm, width = .7\textwidth]{syj_cyclecount2_pit.png}
%		\caption{Cycles that each node in Suyeonjang belongs to. } 
%				\label{fig:10}
%	\end{center}
%\end{figure}

%%%%%%%%%%%%
%%%%%%%%%%%%
%%%%%%%%%%%%%%%%%%%%%%%%%%%%%
%%%%%%%%%%%%%%%%%%%%%%%%%%%%%

\subsection{Songkuyeo}
\begin{figure}[h]
    \begin{center}
    \includegraphics[width=0.6\textwidth]{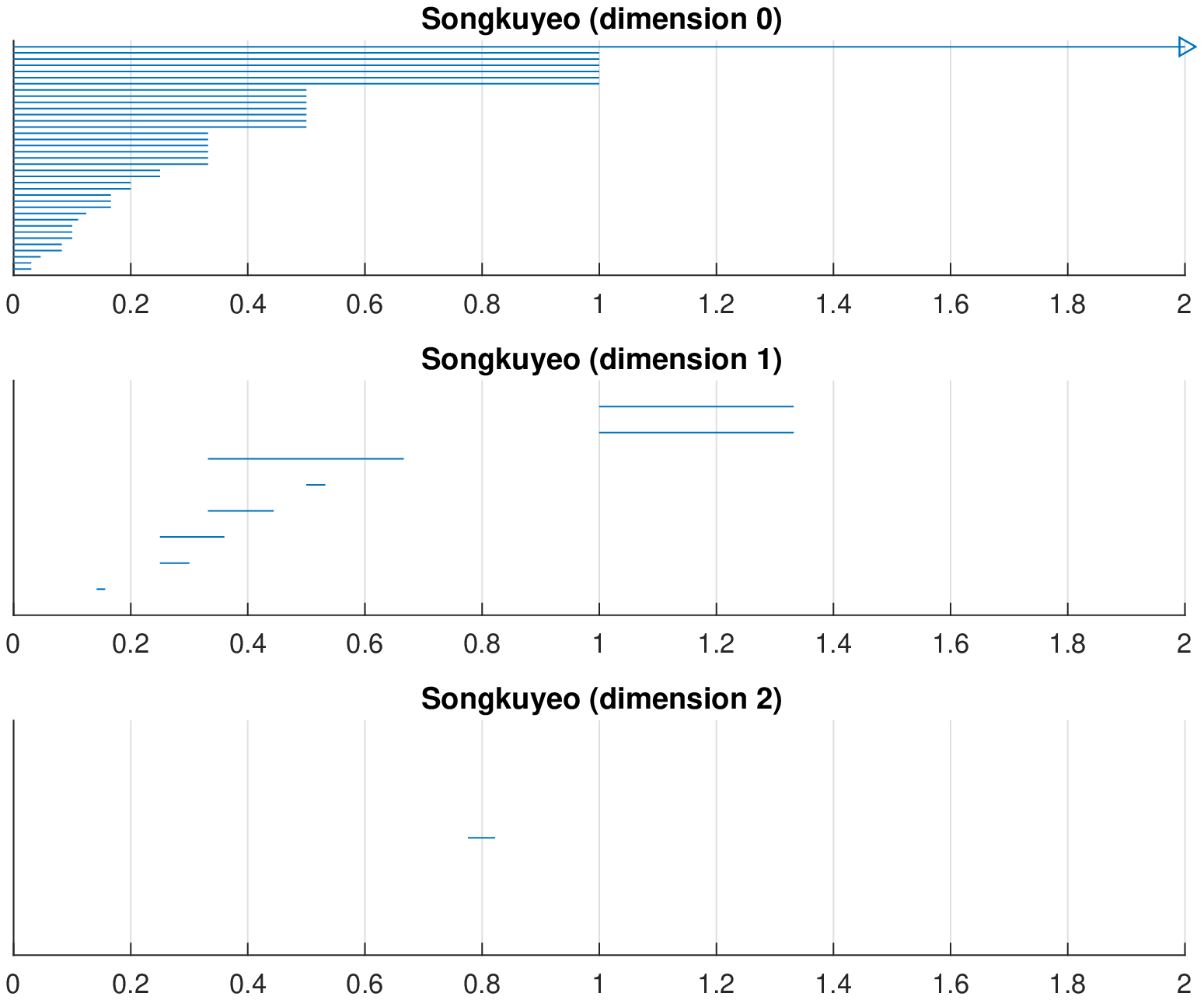}
    \caption{Barcode of Songkuyeo using Vietoris-Rips method for 0-D, 1-D and 2-D. We observe there are $8$ cycles in 1-D barcode while there is one void structure in 2-D barcode. }
    \label{fig:songkuyeo_barcode}
    \end{center}
\end{figure}

We provide the similar data for Songkuyeo. \figref{fig:songkuyeo_barcode} shows the barcodes in 0-D, 1-D and 2D. We observe that there are $37$ components when $\tau = 0$, which corresponds to the total number of nodes in Songkuyeo. In 1D barcode, we observe that there are $8$ non-zero persistences implying that there are $8$ cycles in music network as Suyeonjang. It is interesting to observe that there is one persistence in 2D. This implies that there exists a void-like structure in Songkuyeo network although the size of the void is small as the size of the persistence is small.

% songkuyeo barcode
%%%%%%%%%%%%%%%%%%%%%%

\figref{fig:songkuyeo_cycles} shows the $8$ cycles identified by TDA in Songkuyeo.  Cycle 1 is corresponding to the shortest persistence in 1D barcode. Cycle 6 is corresponding to the longest persistence in 1D barcode. Cycles $2, 3, 5, 6, 7$ and $8$ do not appear in actual music in their whole consecutive form. The persistences of Cycles 7 and 8  start from $\tau = 1$ in 1D barcode. The number of nodes in the cycles in Songkuyeo ranges from $4$ to $8$.  Recall that the maximum number of nodes in a cycle for Suyeonjang is $6$.  The average node number is {\textcolor{black}{$5.375$}} which is {\textcolor{black}{larger}} than the average node number of Suyeonjang. The average weight is {\textcolor{black}{ $7.84125$}}, which is {\textcolor{black}{smaller}} than the average weight of Suyeonjang. 
The information of corresponding music notes is given in Table \ref{tab:songkuyeo_nodeinfo}.

\include{songkuyeo_cycles}
\include{songkuyeo_nodeinfo}
%%%%%%%%%%%%%%%%%%%%%%

%%%%%%%%%%%%%%%%%%%%%%%%%%
% Songkuyeo sequences            
Unlike Suyeonjang, there are only two cycles that appear in the music as a whole consecutive form, Cycle 1 and Cycle 4. 

The following Cycle 1 appears in the music but is not a closed cycle. 
\begin{figure}[h]\begin{center}
            %sequence 1.
            \begin{center}\begin{tikzpicture}[scale = 1]
                \tikzstyle{every node}=[draw,shape=circle,inner sep=0pt,minimum size=.6cm,inner sep=0pt,minimum size=.6cm];
                \node[fill=blue!99] (v1) at (1,0) {\textcolor{white}{林}};%{$3$};
                \node[fill=orange!99] (v2) at (2,0) {\textcolor{white}{南}};%{$19$};
                \node[fill=teal!60] (v3) at (3,0) {\textcolor{black}{林}};%{$5$};
                \node[fill=blue!99] (v4) at (4,0) {\textcolor{white}{潢}};%{$4$};
%                \node[draw=none] (v01) at (1.1,-0.5) {$n_{3}$};
%                \node[draw=none] (v02) at (2.1,-0.5) {$n_{19}$};
%                \node[draw=none] (v03) at (3.1,-0.5) {$n_{5}$};
%                \node[draw=none] (v04) at (4.1,-0.5) {$n_{4}$};
                \node[draw=none] (v01) at (1.1,-0.5) {$n_{20}$};
                \node[draw=none] (v02) at (2.1,-0.5) {$n_{27}$};
                \node[draw=none] (v03) at (3.1,-0.5) {$n_{18}$};
%                \node[draw=none] (v04) at (4.1,-0.5) {$n_{32}$};
                \node[draw=none] (v04) at (4.1,-0.5) {$n_{31}$};
                \draw (v1) -- (v2)
                    (v2) -- (v3)
                    (v3) -- (v4);
            \end{tikzpicture}\end{center}       
                       \end{center}
          % \caption{Sequences found in Songkuyeo}
\end{figure}

The following Cycle 4 also appears in the music and is a closed cycle in terms of pitch. 
\begin{figure}[h]\begin{center}
            %sequence 1.
            %sequence 2.
            \begin{center}\begin{tikzpicture}[scale = 1]
                \tikzstyle{every node}=[draw,shape=circle,inner sep=0pt,minimum size=.6cm,inner sep=0pt,minimum size=.6cm];
                \node[fill=blue!99] (v1) at (1,0) {\textcolor{white}{太}};%{$11$};
                \node[fill=orange!99] (v2) at (2,0) {\textcolor{white}{林}};%{$13$};
                \node[fill=teal!60] (v3) at (3,0) {\textcolor{black}{太}};%{$14$};
                \node[fill=blue!99] (v4) at (4,0) {\textcolor{white}{林}};%{$3$};
%                \node[draw=none] (v01) at (1.1,-0.5) {$n_{11}$};
%                \node[draw=none] (v02) at (2.1,-0.5) {$n_{13}$};
%                \node[draw=none] (v03) at (3.1,-0.5) {$n_{14}$};
%                \node[draw=none] (v04) at (4.1,-0.5) {$n_{3}$};
                \node[draw=none] (v01) at (1.1,-0.5) {$n_{8}$};
                \node[draw=none] (v02) at (2.1,-0.5) {$n_{21}$};
                \node[draw=none] (v03) at (3.1,-0.5) {$n_{6}$};
                \node[draw=none] (v04) at (4.1,-0.5) {$n_{20}$};
                \draw (v1) -- (v2)
                    (v2) -- (v3)
                    (v3) -- (v4);
            \end{tikzpicture}
            \end{center}            
           \end{center}
          % \caption{Sequences found in Songkuyeo}
\end{figure}

\begin{figure}[!h]
	\begin{center}
            	\includegraphics[height = 5cm, width = \textwidth]{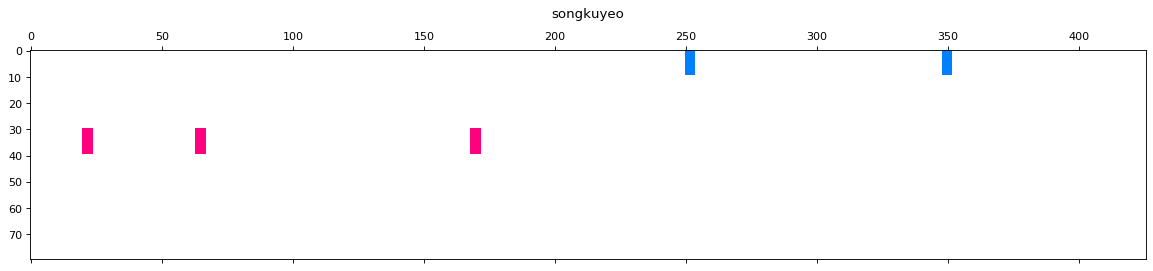}
		\caption{Consecutive sequences of notes corresponding to Cycles 1 and 4 found in Songkuyeo. The horizontal axis is the time sequence of music flow and the vertical axis is for the cycles. The top bars represents Cycle 1 and the bottom Cycle 4. The width of each bar is the number of nodes contained in each cycle. The height of each bar is only for the visual distinction purpose. Note that only two cycles among eight are appearing. } 
		\label{figure12}
	\end{center}
\end{figure}

Figure \ref{figure12} shows the actual cycles appearing in Songkuyeo in their whole consecutive form. As shown in the figure, there are only two cycles, Cycle 1 and Cycle 4 appearing in the music. Cycle 4 appears in the first half of the music and Cycle 1 in the second half of the music. As in Suyeonjang, the distribution of these cycles in the music is sparse. 
%%%

\begin{figure}[!h]
	\begin{center}
            	\includegraphics[height = 4cm, width = \textwidth]{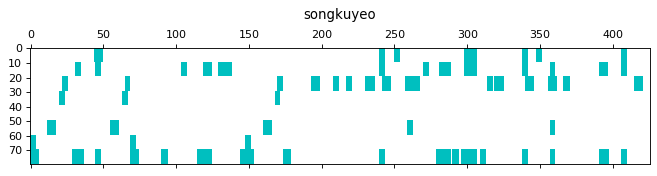}
		\caption{Overlap matrix on $4$-scale for Songkuyeo. This shows consecutive sequences of at least $4$ nodes involved in cycles of Songkuyeo. Cycles are displayed when at least $4$ nodes in the cycle are appearing in the music.  Notice that all cycles except Cycle 5 are appearing in the figure. } 
		\label{fig:sky_min4notes}
	\end{center}
\end{figure}

Figure \ref{fig:sky_min4notes} shows the Overlap matrix on $4$-scale for Songkuyeo. This is the cycle occurrence with respect to music flow  whenever at least $4$ nodes in the cycle are appearing in the music. The order may not be the same for each occurrence. 
%The spectrum shown in the width shows the variation of the order of nodes in the cycle when appearing in the music. 
We observe that the occurrence becomes now  dense. We also observe that all the cycles except Cycle 5 are appearing. It is interesting to notice that Cycle 4 still appears only 3 times.  Also notice again that multiple cycles can be found simultaneously.

%%%
%\begin{figure}[!h]
%	\begin{center}
%            	\includegraphics[height = 7cm, width = .7\textwidth]{songkuyeo_cyclecount1_pit.png}
%		\caption{Number of cycles that each node participates in (Songkuyeo).} 
%	\end{center}
%\end{figure}
%%%%
%\begin{figure}[!h]
%	\begin{center}
%            	\includegraphics[height = 7cm, width = .7\textwidth]{songkuyeo_cyclecount2_pit.png}
%		\caption{Cycles that each node participates in (Songkuyeo).} 
%	\end{center}
%\end{figure}

\begin{figure}[h]
	\begin{center}
	\includegraphics[width=1.0\textwidth]{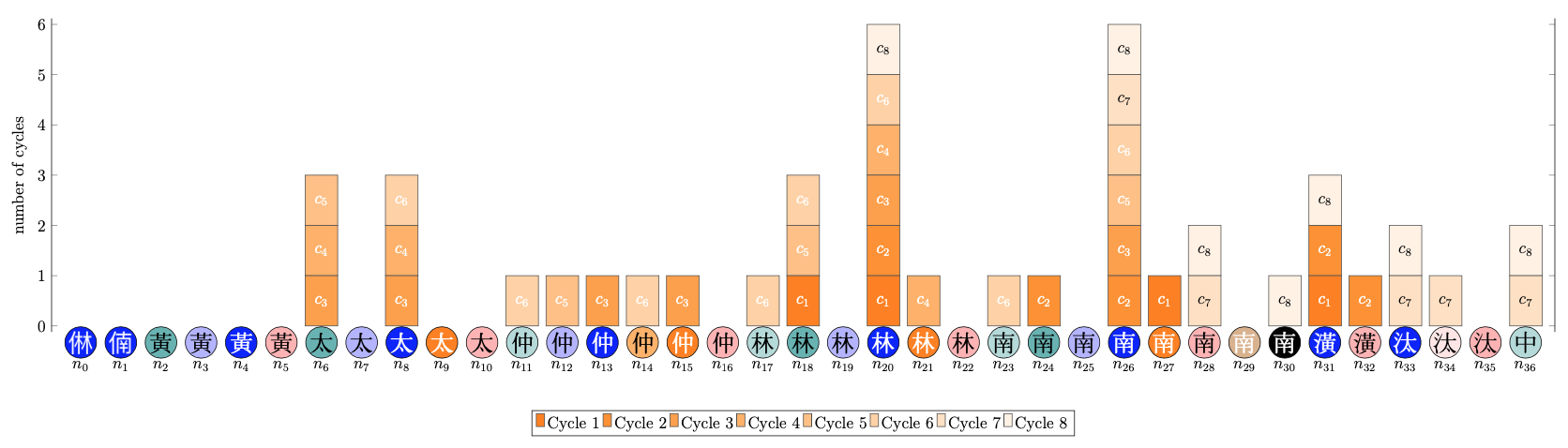}
		\caption{Number of cycles that each node in Songkuyeo belongs to. The horizontal axis shows all nodes in Songkuyeo and the vertical axis shows the total number of cycles that each node belongs to and the actual cycle number.  } 
				\label{fig:sky_numofcyc}
	\end{center}
\end{figure}

Figure \ref{fig:sky_numofcyc} shows how many cycles are there that each of $37$ nodes in Songkuyeo in  \tabref{tab:songkuyeo_allnodes} belongs to. The horizontal axis shows all $37$ nodes in ascending order in terms of pitch and the vertical axis shows the total number of cycles that the node belongs to and cycle number together. The visualization follows the same format of Figure \ref{fig:syj_numofcyc}. 
%
%For examples, each of the first three nodes $n_0, n_1, n_2, n_3$ belongs to only one cycle, i.e.Cycle 4 (c4), Cycle 6 (c6), Cycle 4 (c4) and Cycle 2 (c2) while $n_5$ and $n_6$ do not belong to any of cycles and $n_7$ belongs to $5$ cycles, Cycle 2 (c2), Cycle 4 (c4), Cycle 5 (c5), Cycle 6 (c6) and Cycle 7 (c7). Note that the node number $m_j$ is the node number when all nodes from Suyeonjang, Songkuyeo and Taryong are merged and listed in ascending order in terms of pitch. For example, the node $n_5$ in Suyeonjang becomes the node $m_7$ in the set of all nodes from Suyeonjang, Songkuyeo and Taryong. 

%%%%%%%%%%%%%%%%%%%%%%%%%%%%%%%%%%%%%%%%%%%%%%%%%%%%%%%%%%%%%%%%%%%%%%%%%%%%%%%%
%%%%%%%%%%%%%%%%%%%%%%%%%%%%%%%%%%%%%%%%%%%%%%%%%%%%%%%%%%%%%%%%%%%%%%%%%%%%%%%%

\subsection{Taryong}

\begin{figure}[h]
    \begin{center}
        \includegraphics[width=0.6\textwidth]{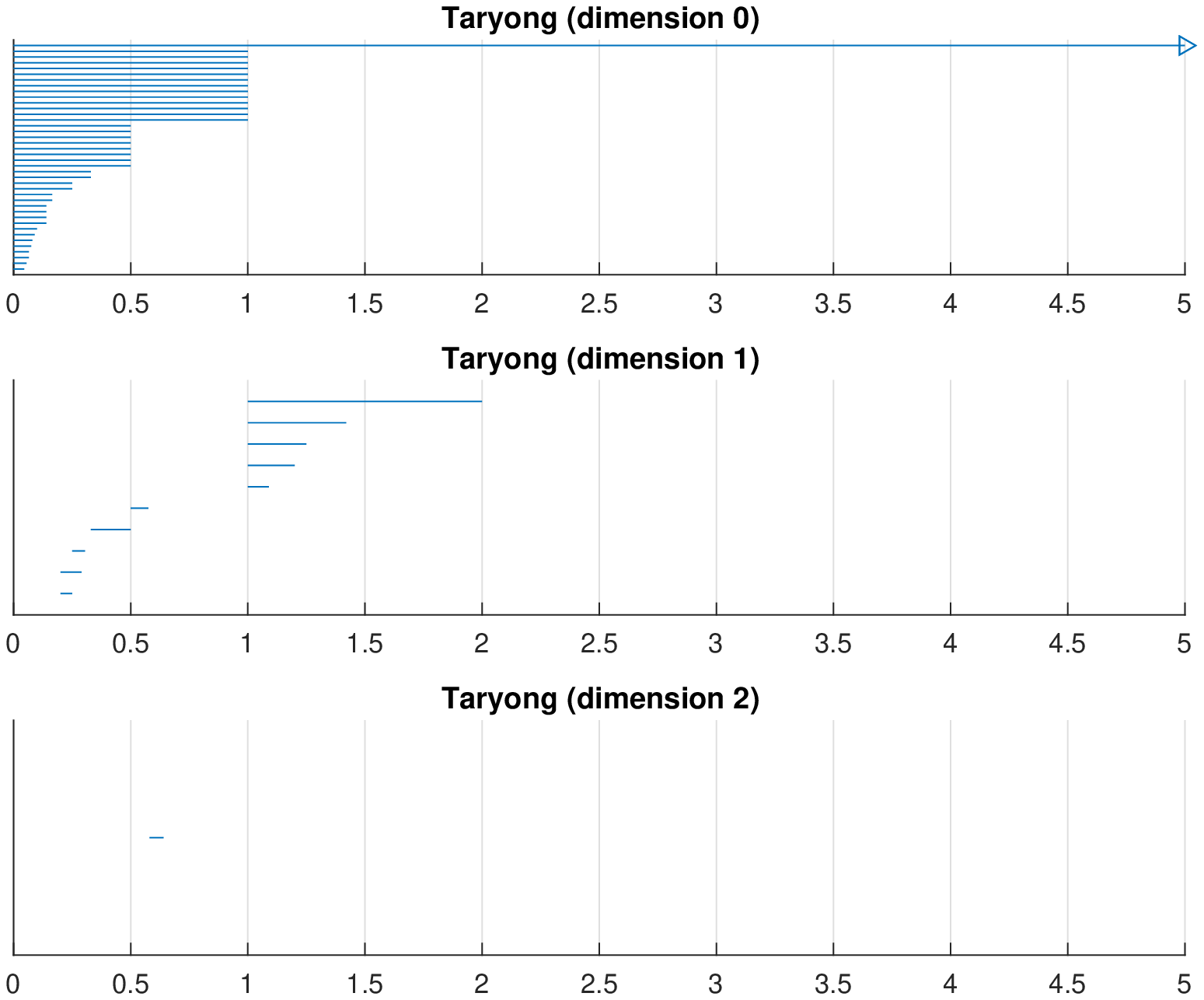}
    \captionof{figure}{Barcode of Taryong using Vietoris-Rips method. Notice that there is a 2D persistence in 2D barcode corresponding to the 2D void in music network. There are $10$ cycles in 1D barcode. }
    \label{fig:taryong_barcode}
    \end{center}
\end{figure}

Figure  \ref{fig:taryong_barcode} shows the barcode of Taryong. As expected, there are $40$ components in 0-D barcode as there are $40$ nodes in Taryong. We observe that there are $10$ cycles in 1D barcode and $1$ cycle (void) in 2-D barcode. The number of 1D cycles of Taryong is larger than both those numbers of Suyeonjang and Songkuyeo.  

The following figure shows the $10$ cycles identified by TDA in Taryong. Each cycle shows {\textcolor{black}{the persistence interval}}, the node number, edge weight {\textcolor{black}{(normal size in blue), distance between node (small size in blue in brackets)}}, and the average weight (red in center). Cycle 1 is corresponding to the shortest persistence in 1D barcode. Cycle 10 is corresponding to the longest persistence in 1D barcode. Cycles $2, 3, 7$ and $8$ do not appear in actual music in their whole consecutive form. The corresponding persistences of Cycles $7$ and $8$ start from $\tau = 1$ while the persistences of Cycles $2$ and $3$ start earlier in 1D barcode. The average node number is $4.8$. The average weight is $4.385$. Notice that the number of cycles identified for Taryong is larger than those numbers of Suyeonjang and Songkuyeo, but the average node number {\textcolor{black}{is smaller than that of Songkuyeo}}. 
%and both the average node}} and average weight are much smaller than those of Suyeonjang and Songkuyeo. 
In fact, despite the larger number of cycles, the occurrence plot of cycles appearing  in the music is much sparser than those plots of Suyeonjang and Songkuyeo, shown in Figures  \ref{figure18} and \ref{fig:tr_min4notes}. 

% taryong barcode

%%%%%%%%%%%%%%%%%%%%%%%%%%%%%
\include{taryong_cycles}         
\include{taryong_nodeinfo} 
%%%%%%%%%%%%%%%%%%%%%%%%%%%%%   

%%%%%%%%%%%%%%%%%%%%%%%%%%%%%
% Taryong sequences
            
\begin{figure}[!h]\centering\resizebox{40ex}{!}{%
\begin{tikzpicture}
\tikzstyle{every node}=[draw,shape=circle,inner sep=0pt,minimum size=.6cm,inner sep=0pt,minimum size=.6cm];
%Sequence 1.
\begin{scope}[shift={(0,0)},scale = 1.3]
            \node[draw=none,rectangle,fontscale=-1] at (2.5,-0.8) {$\text{Cycle } 1$};
            \node[fill=blue!99] (v0) at (0.0,0) {\textcolor{white}{太}};%{$16$};
                \node[draw=none] (n0) at (0.1,-0.4) {$n_{16}$};
                    \node[fill=orange!70] (v1) at (1.0,0) {\textcolor{white}{黃}};%{$12$};
                \node[draw=none] (n1) at (1.1,-0.4) {$n_{12}$};
                    \node[fill=blue!20] (v2) at (2.0,0) {\textcolor{black}{\includegraphics[width=2.35ex]{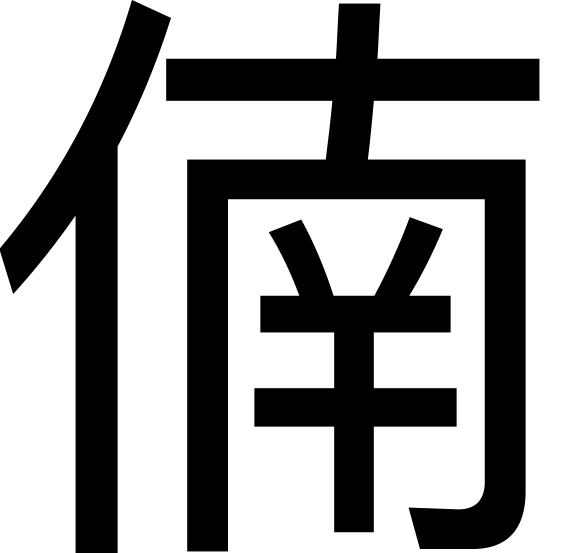}}};%{$6$};
                \node[draw=none] (n2) at (2.1,-0.4) {$n_{6}$};
                    \node[fill=blue!99] (v3) at (3.0,0) {\textcolor{white}{\includegraphics[width=2.35ex]{Imw.png}}};%{$3$};
                \node[draw=none] (n3) at (3.1,-0.4) {$n_{3}$};
                    \node[fill=red!20] (v4) at (4.0,0) {\textcolor{black}{黃}};%{$13$};
                \node[draw=none] (n4) at (4.1,-0.4) {$n_{13}$};
                    \node[fill=blue!99] (v5) at (5.0,0) {\textcolor{white}{太}};%{$16$};
                \node[draw=none] (n5) at (5.1,-0.4) {$n_{16}$};
                \draw (v0) -- (v1);
                \draw (v1) -- (v2);
                \draw (v2) -- (v3);
                \draw (v3) -- (v4);
                \draw (v4) -- (v5);
                \end{scope}
%Sequence 2.
\begin{scope}[shift={(0,-2)},scale = 1.3]
           \node[draw=none,rectangle,fontscale=-1] at (2.5,-0.8) {$\text{Cycle } 4$};
            \node[fill=orange!70] (v0) at (1.0,0) {\textcolor{white}{\includegraphics[width=2.35ex]{Imw.png}}};%{$4$};
                \node[draw=none] (n0) at (1.1,-0.4) {$n_{4}$};
                    \node[fill=blue!20] (v1) at (2.0,0) {\textcolor{black}{\includegraphics[width=2.35ex]{jung0.png}}};%{$0$};
                \node[draw=none] (n1) at (2.1,-0.4) {$n_{0}$};
                    \node[fill=blue!99] (v2) at (3.0,0) {\textcolor{white}{\includegraphics[width=2.35ex]{Imw.png}}};%{$3$};
                \node[draw=none] (n2) at (3.1,-0.4) {$n_{3}$};
                    \node[fill=red!20] (v3) at (4.0,0) {\textcolor{black}{黃}};%{$13$};
                \node[draw=none] (n3) at (4.1,-0.4) {$n_{13}$};
                \draw (v0) -- (v1);
                \draw (v1) -- (v2);
                \draw (v2) -- (v3);
                \end{scope}
%Sequence 3.
\begin{scope}[shift={(0,-4)},scale = 1.3]
            \node[draw=none,rectangle,fontscale=-1] at (2.5,-0.8) {$\text{Cycle } 5$};
            \node[fill=orange!70] (v0) at (1.0,0) {\textcolor{white}{黃}};%{$12$};
                \node[draw=none] (n0) at (1.1,-0.4) {$n_{12}$};
                    \node[fill=blue!20] (v1) at (2.0,0) {\textcolor{black}{\includegraphics[width=2.35ex]{Im0.png}}};%{$2$};
                \node[draw=none] (n1) at (2.1,-0.4) {$n_{2}$};
                    \node[fill=blue!99] (v2) at (3.0,0) {\textcolor{white}{黃}};%{$11$};
                \node[draw=none] (n2) at (3.1,-0.4) {$n_{11}$};
                    \node[fill=red!20] (v3) at (4.0,0) {\textcolor{black}{太}};%{$18$};
                \node[draw=none] (n3) at (4.1,-0.4) {$n_{18}$};
                \draw (v0) -- (v1);
                \draw (v1) -- (v2);
                \draw (v2) -- (v3);
                \end{scope}
%Sequence 4.
\begin{scope}[shift={(0,-6)},scale = 1.3]
            \node[draw=none,rectangle,fontscale=-1] at (2.5,-0.8) {$\text{Cycle } 6$};
            \node[fill=blue!99] (v0) at (0.5,0) {\textcolor{white}{黃}};%{$11$};
                \node[draw=none] (n0) at (0.6,-0.4) {$n_{11}$};
                    \node[fill=brown!20] (v1) at (1.5,0) {\textcolor{black}{太}};%{$19$};
                \node[draw=none] (n1) at (1.6,-0.4) {$n_{19}$};
                    \node[fill=red!20] (v2) at (2.5,0) {\textcolor{black}{仲}};%{$24$};
%                \node[draw=none] (n2) at (2.6,-0.4) {$n_{24}$};
                \node[draw=none] (n2) at (2.6,-0.4) {$n_{23}$};
                    \node[fill=blue!99] (v3) at (3.5,0) {\textcolor{white}{太}};%{$16$};
                \node[draw=none] (n3) at (3.6,-0.4) {$n_{16}$};
                    \node[fill=red!20] (v4) at (4.5,0) {\textcolor{black}{林}};%{$30$};
%                \node[draw=none] (n4) at (4.6,-0.4) {$n_{30}$};
                \node[draw=none] (n4) at (4.6,-0.4) {$n_{29}$};
                \draw (v0) -- (v1);
                \draw (v1) -- (v2);
                \draw (v2) -- (v3);
                \draw (v3) -- (v4);
                \end{scope}
%Sequence 5.
\begin{scope}[shift={(0,-8)},scale = 1.3]
            \node[draw=none,rectangle,fontscale=-1] at (2.5,-0.8) {$\text{Cycle } 9$};
            \node[fill=red!20] (v0) at (-0.5,0) {\textcolor{black}{林}};%{$30$};
%                \node[draw=none] (n0) at (-0.4,-0.4) {$n_{30}$};
                \node[draw=none] (n0) at (-0.4,-0.4) {$n_{29}$};
                    \node[fill=blue!99] (v1) at (0.5,0) {\textcolor{white}{林}};%{$27$};
%                \node[draw=none] (n1) at (0.6,-0.4) {$n_{27}$};
                \node[draw=none] (n1) at (0.6,-0.4) {$n_{26}$};
                    \node[fill=red!20] (v2) at (1.5,0) {\textcolor{black}{南}};%{$34$};
%                \node[draw=none] (n2) at (1.6,-0.4) {$n_{34}$};
                \node[draw=none] (n2) at (1.6,-0.4) {$n_{33}$};
%                    \node[fill=red!40] (v3) at (2.5,0) {\textcolor{black}{潢}};%{$38$};
                    \node[fill=red!90] (v3) at (2.5,0) {\textcolor{white}{潢}};%{$38$};
%                \node[draw=none] (n3) at (2.6,-0.4) {$n_{38}$};
                \node[draw=none] (n3) at (2.6,-0.4) {$n_{37}$};
%                    \node[fill=blue!70] (v4) at (3.5,0) {\textcolor{white}{南}};%{$32$};
                    \node[fill=teal!30] (v4) at (3.5,0) {\textcolor{black}{汰}};%{$32$};
%                \node[draw=none] (n4) at (3.6,-0.4) {$n_{32}$};
                \node[draw=none] (n4) at (3.6,-0.4) {$n_{38}$};
                    \node[fill=blue!99] (v5) at (4.5,0) {\textcolor{white}{南}};%{$33$};
%                \node[draw=none] (n5) at (4.6,-0.4) {$n_{33}$};
                \node[draw=none] (n5) at (4.6,-0.4) {$n_{32}$};
                    \node[fill=red!20] (v6) at (5.5,0) {\textcolor{black}{林}};%{$30$};
%                \node[draw=none] (n6) at (5.6,-0.4) {$n_{30}$};
                \node[draw=none] (n6) at (5.6,-0.4) {$n_{29}$};
                \draw (v0) -- (v1);
                \draw (v1) -- (v2);
                \draw (v2) -- (v3);
                \draw (v3) -- (v4);
                \draw (v4) -- (v5);
                \draw (v5) -- (v6);
                \end{scope}
%Sequence 6.
\begin{scope}[shift={(0,-10)},scale = 1.3]
            \node[draw=none,rectangle,fontscale=-1] at (2.5,-0.8) {$\text{Cycle } 10$};
            \node[fill=blue!20] (v0) at (0.5,0) {\textcolor{black}{黃}};%{$10$};
                \node[draw=none] (n0) at (0.6,-0.4) {$n_{10}$};
                    \node[fill=red!70] (v1) at (1.5,0) {\textcolor{white}{仲}};%{$25$};
%                \node[draw=none] (n1) at (1.6,-0.4) {$n_{25}$};
                \node[draw=none] (n1) at (1.6,-0.4) {$n_{24}$};
                    \node[fill=red!20] (v2) at (2.5,0) {\textcolor{black}{林}};%{$30$};
%                \node[draw=none] (n2) at (2.6,-0.4) {$n_{30}$};
                \node[draw=none] (n2) at (2.6,-0.4) {$n_{29}$};
                    \node[fill=blue!99] (v3) at (3.5,0) {\textcolor{white}{仲}};%{$22$};
%                \node[draw=none] (n3) at (3.6,-0.4) {$n_{22}$};
                \node[draw=none] (n3) at (3.6,-0.4) {$n_{21}$};
                    \node[fill=red!20] (v4) at (4.5,0) {\textcolor{black}{太}};%{$18$};
                \node[draw=none] (n4) at (4.6,-0.4) {$n_{18}$};
                \draw (v0) -- (v1);
                \draw (v1) -- (v2);
                \draw (v2) -- (v3);
                \draw (v3) -- (v4);
                \end{scope}
\end{tikzpicture}}
	\caption{Cycles appearing in Taryong as a whole consecutive form.}
	\label{fig:TRseq}
\end{figure}
\begin{figure}[!h]
	\begin{center}
            	\includegraphics[height = 5cm, width = \textwidth]{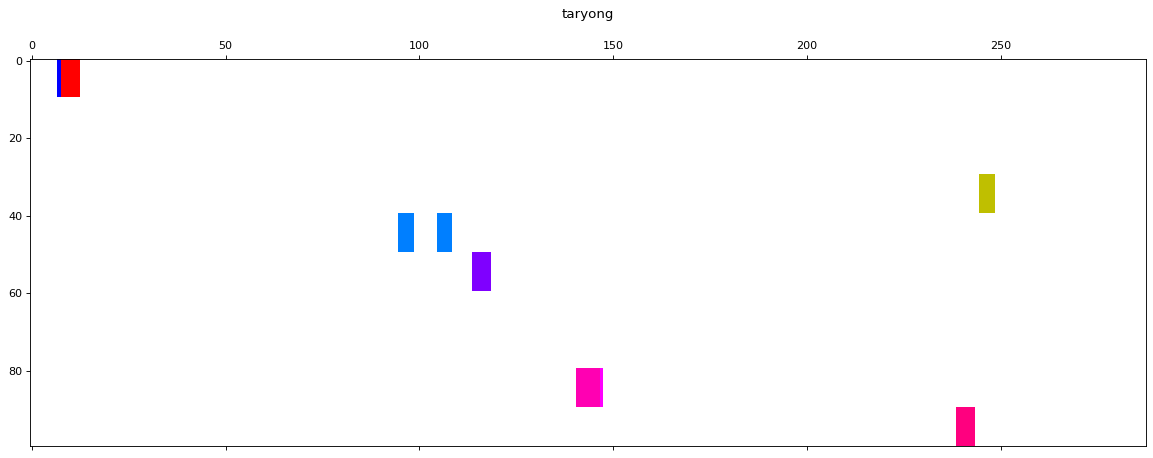}
		\caption{Consecutive sequences of nodes corresponding to Cycles 1, 4, 5, 6,  9 and 10 found in Taryong. The horizontal axis is the time sequence of music flow and the vertical axis is for the cycles. The top bars represent Cycle 1 and the bottom Cycle 10. The width of each bar is the number of nodes contained in each cycle. The height of each bar is only for the visual distinction purpose. Note that {\textcolor{black} {six cycles among all cycles}} are appearing in the music.} 
		\label{figure18}
	\end{center}
\end{figure}
%%%%%%%%%%%%%%%%%%%%%%%%%%

{\textcolor{black}{Figures \ref{fig:TRseq} and}} \ref{figure18} show the consecutive sequences of nodes corresponding to Cycles 1, 4, 5, 6,  9 and 10 found in Taryong. {\textcolor{black}{Among them Cycles 1 and  9 are closed cycles.} 
%The horizontal axis of \figref{figure18} is the time sequence of music flow and the vertical axis is for the cycles. 
The top bars represent Cycle 1 and the bottom  bars represent Cycle 10.
 %The width of each bar is the number of nodes contained in each cycle. The height of each bar is only for the visual distinction purpose. 
 Note that {\textcolor{black} {six cycles among all cycles}} are appearing in the music.  As Suyeonjang and Songkuyeo, the occurrence distribution is sparse.  

%%%
\begin{figure}[!h]
	\begin{center}
            	\includegraphics[height = 5cm, width = \textwidth]{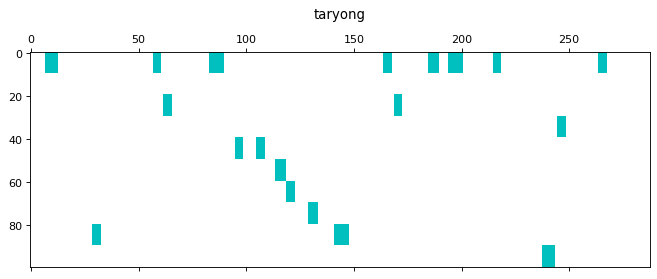}
		\caption{Overlap matrix on $4$-scale for Taryong. This shows consecutive sequences of at least $4$ nodes involved in cycles of Taryong. Cycles are displayed when at least $4$ nodes in the cycle are appearing in the music.  Notice that all cycles except Cycle 2 are appearing in the figure. } 
		\label{fig:tr_min4notes}
	\end{center}
\end{figure}

Figure \ref{fig:tr_min4notes} shows the Overlap matrix on $4$-scale for Taryong. This shows consecutive sequences of at least $4$ nodes involved in cycles of Taryong. Cycles are displayed when at least $4$ nodes in the cycle are appearing in the music.  Notice that all cycles except Cycle 2 are appearing in the figure. It is interesting to observe that unlike Suyeonjang and {\textcolor{black}{Songkuyeo}} the plot is still spare although Taryong has more cycles than Suyeonjang and Songkuyeo. 

\textcolor{black}{We compare Figures \ref{fig:syj_min4notes}, \ref{fig:sky_min4notes} and \ref{fig:tr_min4notes} in Table \ref{tab:comparison}. 
The denseness is defined as follows
\[
	\text{Denseness} = \frac{A_c}{A_f},
\]
where $A_c$ is the total area that consecutive sequences of at least 4 nodes belonging to the cycles occupy, and $A_f$ is the total area of the corresponding Figure.
}
We see that  the denseness of cycles of Suyeonjang and Songkuyeo are about three times larger than that of Taryong. %Here, the denseness is defined as the ratio between the total area that consecutive sequences of at least 4 nodes belonging to the cycles occupy and the total area of the corresponding Figure.
\textcolor{black}{It is also observed that, unlike Suyeonjang and Songkuyeo, the consecutive sequences of at least 4 nodes involved in cycles of Taryong are non-overlapping, meaning that only one cycle occurs at a time as music flows. So, if we define the overlapping percentage by% the ratio between the number of times at least two cycles occurred simultaneously and the number of times a cycle occurred in the time sequence of music flow, 
\[
	\text{Overlap} = \frac{N_s}{N_c} \times 100\%,
\]
where $N_s$ is the number of times at least two cycles occurred simultaneously, and $N_c$ is the number of times a cycle occurred in the time sequence of music flow, 
then the consecutive sequences of at least 4 nodes involved in cycles of Suyeonjang and Songkuyeo overlap $36.2318\%$ and $32.05128\%$, respectively, and those of Taryong overlap $0\%$. The results are showed in the last column of Table \ref{tab:comparison}. It is reasonable to hypothesize that the cycle overlapping gives Suyeonjang and Songkuyeo a feeling of melodic repetition which is well matched with the fact that they are classified as {\color{black}{Dodeuri}}, literally translated as "repetition", in Korean traditional music.
}

\begin{table}[!h]
  \begin{center}
    \caption{Comparison of cycle distributions}
    \label{tab:comparison}
    \resizebox{\columnwidth}{!}{%
    \begin{tabular}{| c || c | c | c| c| c| c|} % <-- Alignments: 1st column left, 2nd middle and 3rd right, with vertical lines in between
    \hline
         & \# of cycles & Average node \# & Average weight & Occurrence/Cycle   & Denseness    & Overlap  (\%)  \\
      \hline
      Suyeonjang & 8 &  {\textcolor{black}{4.625}} & {\textcolor{black}{9.39375}} & 1.125 & 0.09489 & 36.23188 \\
      Songkuyeo & 8 &  {\textcolor{black}{5.375}} & {\textcolor{black}{7.84125}} & 0.625 & 0.10153 & 32.05128 \\
      Taryong & 10 & 4.8 & 4.385 & 0.7 &  0.03194 & 0\\
      \hline
    \end{tabular}%
    }
  \end{center}
\end{table}

%%%%%%%%
\begin{figure}[!h]
	\begin{center}
	\includegraphics[width=1.0\textwidth]{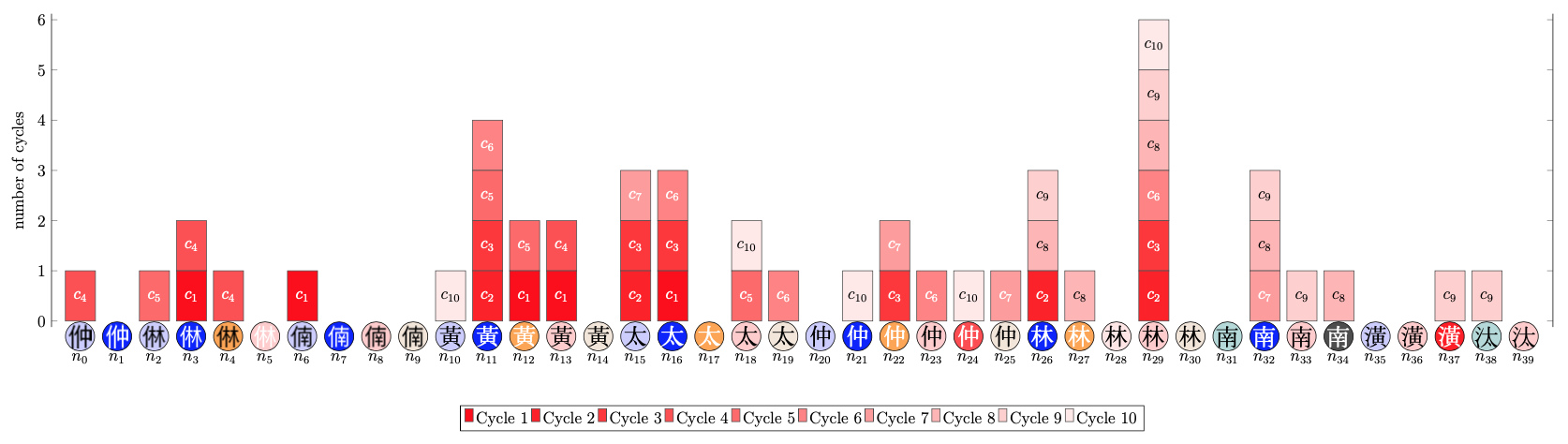}
		\caption{Number of cycles that each node in Taryong belongs to. The horizontal axis shows all nodes in Taryong and the vertical axis shows the total number of cycles that each node belongs to and the actual cycle number.  } 
				\label{fig:tr_numofcyc}
	\end{center}
\end{figure}

Figure \ref{fig:tr_numofcyc} shows how many cycles are there that each of $40$ nodes in Taryong in  \tabref{tab:taryong_allnodes} belongs to. The horizontal axis shows all $40$ nodes in ascending order in terms of pitch and the vertical axis shows the total number of cycles that the node belongs to and the corresponding cycle number together. The visualization follows the same format of Figure \ref{fig:syj_numofcyc}. 

\section{Concluding remark}
In this work we used persistent homology to identify the cycles in three famous old Korean music pieces, namely Suyeonjang, Songkuyeo and Taryong by constructing the simplicial complex of music network of those three pieces.  By calculating homology at different scales, we found several one-dimensional cycles in each music. By visualizing the distribution of the cycles in the music flow, we showed how those identified cycles appear through the music flow. We found that the first two music pieces, Suyeonjang and Songkuyeo, show similar overlapping patterns of those cycles in a way that cycles appear simultaneously while Taryong shows different patterns that cycles only appear individually. Our finding provides a systematic explanation of the unique structure embedded in Suyeonjang and Songkuyeo known as {\color{black}{"Dodeuri"}} (repeat-and-return) pattern. That is, {\color{black}{"Dodeuri"}} pattern does not only repeat or variate themes but also juxtapose those parallely to maximize its musical inspiration of cyclic patterns. In this paper, we showed that persistent homology provides a useful tool to analyze the Korean music written in Jeongganbo. Our current work did not consider the ornamenting tones and the analysis was based on the instrument-specific notes. In our future work, we plan to consider the ornamenting tones and expand the analysis to other instruments using persistent homology.

\vskip .2in
\noindent
{\bf{Acknowledgements:} } This work is supported by the NRF SGER grant under grant number 2021R1A2C33009648. This research was partially supported by Samsung Science \& Technology Foundation SSTF-BA1802-02

%%%%%%%%%%%%
%%%%%%%%%%%%
%\appendix
%\section*{Appendix}
%%%%
%\renewcommand{\thesubsection}{\Alph{subsection}}
%\subsection{All nodes in Suyeonjang, Taryong, Songkuyeo}
%\include{syj_allnodes}
%\include{songkuyeo_allnodes}
%\include{taryong_allnodes}
%%%
%\subsection{If we merge the node list of three musics then indexing}
%\include{allresult_merged}
%\includepdf[page=-]{cyclenumANDfrequency.pdf}

%%%%%%%%%%%%
%%%%%%%%%%%%
%\bibliographystyle{tMAM}
%\bibliography{paper_ref}

\bibliographystyle{plain}
\bibliography{paper_ref}	
%%%%%%%%%%%%
%%%%%%%%%%%%

%\appendices
\section*{Appendix: All nodes in Suyeonjang, Taryong, Songkuyeo}
%%%%
%\renewcommand{\thesubsection}{\Alph{subsection}}
%\section{All nodes in Suyeonjang, Taryong, Songkuyeo}
\include{syj_allnodes}
\include{songkuyeo_allnodes}
\include{taryong_allnodes}
%%%
\end{document}